\UseRawInputEncoding

\documentclass[aps,prmaterials, twocolumn, superscriptaddress, amsmath, amssymb]{revtex4}

\usepackage{graphicx}
\usepackage{dcolumn}
\usepackage{bm}
\usepackage{caption}
\usepackage{layouts}
\usepackage{xcolor}

\begin{document}

\title{Tight-binding bond parameters for dimers across the periodic table from density-functional theory}

\author{Jan Jenke}
\affiliation{ICAMS, Ruhr University Bochum, D-44801 Bochum, Germany}
\author{Alvin N. Ladines}
\affiliation{ICAMS, Ruhr University Bochum, D-44801 Bochum, Germany}
\author{Thomas Hammerschmidt}
\affiliation{ICAMS, Ruhr University Bochum, D-44801 Bochum, Germany}
\author{David G. Pettifor}\affiliation{University of Oxford, Oxford OX1-3PH, United-Kingdom}
\author{Ralf Drautz}
\affiliation{ICAMS, Ruhr University Bochum, D-44801 Bochum, Germany}

\date{\today}

\begin{abstract}
We obtain parameters for non-orthogonal and orthogonal TB models from two-atomic molecules for all combinations of elements of period 1 to 6 and group 3 to 18 of the periodic table. The TB bond parameters for 1711 homoatomic and heteroatomic dimers show clear chemical trends. In particular, using our parameters we compare to the rectangular $d$-band model, the reduced $sp$ TB model as well as canonical TB models for $sp$- and $d$-valent systems which have long been used to gain qualitative insight into the interatomic bond. The transferability of our dimer-based TB bond parameters to bulk systems is discussed exemplarily for the bulk ground-state structures of Mo and Si. 
Our dimer-based TB bond parameters provide a well-defined and promising starting point for developing refined TB parameterizations and for making the insight of TB available for guiding materials design across the periodic table.
\end{abstract}

\pacs{}

\maketitle

\section{\label{sec:introduction}Introduction}

The discovery and design of materials requires the prediction of structural and functional properties for a given atomic structure and chemical composition. 
Density functional theory (DFT) enables accurate quantum mechanical simulations of the interatomic interaction and therefore became a standard tool in materials science. 
As the computational effort of DFT calculations is considerable and rises rapidly with system size, systematic searches for desired materials properties are often limited to small subsets of atomic structures and chemical compositions. Coarser models are required for searching large parameter spaces faster than DFT. These may be obtained from data analysis of DFT or experimental data sets (see e.g. Refs.~\cite{0022-3719-19-3-002,doi:10.1021/acs.chemmater.5b04299}) or a direct simplification of DFT. The tight-binding (TB) bond model~\cite{0022-3719-21-1-007,PhysRevB.84.214114,0965-0393-23-7-074004} is derived from a systematic coarse-graining of DFT by a second order expansion of the DFT functional with respect to the charge density and provides a robust and physically intuitive description of the interatomic bond. 

TB models can be divided largely in two complementary groups, broad models to rationalize chemical and structural trends and specific parameterizations for modeling particular materials. Prominent representatives of the first group are the rectangular $d$-band model~\cite{PETTIFOR198743} and canonical TB models~\cite{PhysRevB.17.1209,harrison,Turchi-91,0953-8984-3-5-001} that have been successfully applied for qualitative analysis (see e.g. Refs.~\cite{0953-8984-3-5-001, PhysRevLett.53.1080, PhysRevB.72.144105, Hammerschmidt-08-2, PhysRevB.83.224116}) and as the basis of machine-learning descriptors~\cite{Jenke-18,Sutton-19}, but cannot be employed for quantitative predictions. 
Examples of the second group include parameterizations of the Naval Research Laboratory tight-binding (NRL-TB) formalism~\cite{0953-8984-15-10-201}, density functional tight-binding (DFTB)~\cite{PhysRevB.51.12947, PhysRevB.58.7260}, or the geometry, frequency, noncovalent, extended TB (GFN-xTB) method~\cite{doi:10.1021/acs.jctc.7b00118}.

We parameterize the TB bond parameters for homoatomic and heteroatomic dimers, i.e. diatomic molecules, across the periodic table. 
Dimers have been extensively studied in numerous experimental works and many properties are tabulated~\cite{HuberHerzberg}. The theoretical treatment of dimers by electronic-structure calculations can be challenging, see e.g. Ref.~\cite{Lehtola-19} for a recent review. 
Wave-function based quantum-chemistry methods that are required for highly accurate predictions of certain dimers (see e.g. Ref.~\cite{Brynda-09} for Cr-Cr) are too computationally expensive to cover large parts of the periodic table. 
Density-based methods like DFT have known shortcomings in describing dimers~\cite{Gunnarsson-85,Kurth-99,Ernzerhof-99,Barden-00,Gutsev-03} depending on the exchange-correlation (XC) functional. 
As in other works on large sets of molecules~\cite{Chaves-17}, we use the PBE  functional without further refinements in order to treat all dimers on the equal footing of a non-empirical XC functional. From the close connection to physical principles in the construction of the PBE functional, see e.g. Ref.~\cite{Ernzerhof-99,Kurth-99}, we expect a robust treatment of chemical trends in the periodic table.

For the parameterization we apply a downfolding procedure~\cite{PhysRevB.83.184119, PhysRevB.84.155119} for dimers of all combinations of elements of period 1 to 6 and group 3 to 18. In this way we establish a database of pairwise interaction parameters as starting point for the parameterization of TB models across the periodic table.
In Sec.~\ref{sec:methodology} we summarize the procedure for downfolding the DFT wavefunction and for parameterizing the TB models. In Sec.~\ref{sec:Si_analysis} this is illustrated for the Si and Mo dimers. Trends across the elements and parameterizations are examined for homoatomic dimers in Sec.~\ref{sec:homo} and for heteroatomic dimers in Sec.~\ref{sec:hetero}. In Sec.~\ref{sec:compare} we compare our parameterizations to available TB models. After a brief discussion of the transferability to bulk materials in Sec.~\ref{sec:bulk}, we conclude in Sec.~\ref{sec:conclusion}.

\section{TB parameterization\label{sec:methodology}}

The development of a TB bond model starts from a pairwise parameterization of the Hamiltonian matrix elements. 
In two-center approximation the number of independent matrix elements or bond integrals is significantly reduced by taking into account rotational invariance~\cite{PhysRev.94.1498}. 
For a dimer in two-center approximation, we determined the Hamiltonian matrix in bond direction analytically as
\begin{equation}\label{eq:Hdimer}
H = 
\begin{pmatrix}
\sigma_0&	0&			0&			0&				0\\
0&			\pi_{-1}	&	0&			0&				0\\
0&			0&			\pi_{+1}	&	0&				0\\
0&			0&			0&			\delta_{-2}&		0\\
0&			0&			0&			0&				\delta_{+2}\\
\end{pmatrix},
\end{equation}
where $\sigma_0$, $\pi_{\pm 1}$ and $\delta_{\pm 2}$ are block matrices. The required matrix elements are given in appendix~\ref{app:BlockMatrices}. We determine the numerical values of the matrix elements by employing a downfolding procedure that creates an optimized minimal basis from a multiple-$\zeta$ linear
combination of atomic orbital (LCAO) basis as developed~\cite{PhysRevB.83.184119, PhysRevB.84.155119} and applied~\cite{LADINES20171315, PhysRevB.93.155203, 0953-8984-25-11-115502, PhysRevB.86.155115, 0953-8984-23-27-276004} recently. 

For our database of TB bond parameters we consider all elements from group 3 to group 18 in period 1 to 6 using the projector augmented wave (PAW)~\cite{PhysRevB.50.17953} datasets of GPAW~\cite{PhysRevB.71.035109, 0953-8984-22-25-253202} (setup version 0.9.11271) except of Po, At, Tc and Lu, which were not available from GPAW. 
We assign the elements in group 3 to 11 an $sp$ valence and the elements in group 12 to 18 an $sd$ valence. Hydrogen and helium we consider as $s$-valent.

The TB matrix elements are then parameterized. All matrix elements are represented by a common functional  form that is able to capture the details of the distance dependence for all the 1711 dimers that we considered.
The parameterization of 8476 interatomic matrix elements for each TB matrix and 11310 onsite matrix elements for both, the orthogonal and non-orthogonal TB Hamiltonian matrix leads to 48048 matrix elements that are compiled in the Supplemental Material at Ref.~\cite{SuppMat}.

\subsection{Downfolding\label{sec:downfolding}}

For downfolding the DFT wavefunction to a TB minimal basis we need to choose the DFT reference state. A self-consistent DFT wavefunction mixes effects of the self-consistency from DFT into the TB Hamiltonian.
This is undesirable as self-consistency should not affect the TB Hamiltonian. Therefore, we take the Harris-Foulkes (HF) approximation to DFT~\cite{PhysRevB.31.1770, PhysRevB.39.12520} that is constructed from the electron density $\rho^{(0)}$ of overlapping free atoms as the reference state. The DFT Hamiltonian in HF approximation is given by
\begin{equation}
\begin{split}
\hat{H} = & -\frac{\hbar^2}{2m}\nabla^2 + V_{\mathrm{ion}}(\mathbf{r}) + V^\text{H}[\rho^{(0)}(\mathbf{r})] \\
& + V^{\text{XC}}[\rho^{(0)}(\mathbf{r}), \nabla \rho^{(0)}(\mathbf{r})],
\end{split}
\end{equation}
with the ionic potential $V_{\mathrm{ion}}$, the Hartree potential $V^{\mathrm{H}}$, and the exchange-correlation potential $V^{\mathrm{XC}}$. 

The HF approximation leads to an error in the ground-state total energy that is of second order in the difference between input charge density and self-consistent charge density. Numerous tests found small resulting errors for properly chosen neutral-atom densities and demonstrated a reliable prediction of the HF approximation for total energies of homoatomic~\cite{PhysRevB.31.1770} and heteroatomic dimers~\cite{PhysRevB.39.12520, Averill-90}, molecules~\cite{Bellchambers-11}, solids~\cite{Polatoglou-90, Paxton-90, Finnis-90, Farid-93, NguyenManh-07, Andritsos-19}, and (with additional optimisation of the input density) also surfaces~\cite{Read-90, Finnis-90, Chetty-91}.

The HF-DFT reference states of the dimers are created for interatomic distances from $\min \left( 3/4 R_{\text{eq}}, 3 \text{\AA} \right)$ to $8 \text{\AA}$ in steps of $0.05 \text{\AA}$.
The equilibrium bond length $R_{\text{eq}}$ is calculated for the homoatomic dimers by self-consistent DFT calculations and for heteroatomic dimers approximated by averaging the values of the homoatomic dimers.
The calculations are carried out using GPAW with the Perdew-Burke-Ernzerhof (PBE) exchange-correlation functional~\cite{PhysRevLett.77.3865} and PAW~\cite{PhysRevB.50.17953} datasets. 
For all calculations a constant grid spacing of $h = 0.15 \mathrm{\AA}$ is used. 
We used a non-spinpolarized grid basis in order to ensure that the effects of magnetism are not mixed into an initially non-self-consistent TB Hamiltonian but rather arise from self-consistency at the TB level in combination with, e.g., a Stoner model (see e.g. Ref.~\cite{ford_2014}).

For each bond distance of the dimers we apply a downfolding procedure~\cite{PhysRevB.83.184119} that starts by expanding the HF-DFT eigenstates $| \psi_n \rangle$ in a triple-$\zeta$ basis $\{| \phi_{Ilmz}\rangle\}$,
\begin{equation}
| \psi_n \rangle = \sum_{Ilmz} c_{Ilmz}^{(n)} | \phi_{Ilmz}\rangle \,,
\end{equation}
where $I$ labels the atom, $lm$ the angular character and $z$ the number of radial functions per orbital. 
A TB minimal basis with only one radial function per orbital is obtained from a linear combination of the triple-$\zeta$ basis functions for each angular character
\begin{equation}
|\varphi_{I lm}\rangle = \sum_z k_{Ilmz} |\phi_{Ilmz}\rangle \,.
\end{equation}
This is achieved by numerical optimization of the coefficients $k_{Ilmz}$ that maximize the projection
\begin{equation}
P = \frac{1}{N_e} \sum_n f_n \langle \psi_n | \hat{P} | \psi_n  \rangle \,,
\label{eq:max_P}
\end{equation}
with occupation numbers $f_n$, number of valence electrons $N_e$ and the projection operator
\begin{equation}
\hat{P} = \sum_{I lm} | \varphi_{I lm} \rangle \langle \varphi^{I lm} |\,.
\end{equation}

In general we carried out the downfolding to the basis according to the PAW basis of GPAW. In the following we limit the basis to $s$, $sp$ or $sd$ matrix elements.
To remove the excess orbitals, we separate the optimal minimal basis into basis functions to be included (TB) and not to be included (omit) in the final TB model
\begin{equation}
\{ |\varphi_{Ilm}\rangle\} = \left\{ \{| \varphi_{Ilm}^{\text{TB}}\rangle\}, \{ |\varphi_{Il'm'}^{\text{omit}}\rangle \} \right\}.
\end{equation}
The optimal eigenstates in the reduced basis 
\begin{equation}
|\psi_n^{\text{TB}} \rangle = \sum_{Ilm} \frac{1}{||\psi_n^{\text{TB}}||^{1/2}}c_{Ilm}^{(n), \text{TB}} | \varphi_{Ilm}^{\text{TB}}\rangle \,,
\end{equation}
are determined such that the relevant eigenvalues of the minimal basis Hamiltonian are reproduced while the change in the eigenstates is kept minimal.

The TB Hamiltonian and the TB overlap is then computed as 
\begin{equation}
\begin{split}
H_{IlmI'l'm'}^{\text{TB}} & = \langle \varphi_{Ilm}^{\text{TB}} | \hat{H}^{\text{TB}} |\varphi_{I'l'm'}^{\text{TB}} \rangle \,,\\
S_{IlmI'l'm'}^{\text{TB}} & = \langle \varphi_{Ilm}^{\text{TB}} | \varphi_{I'l'm'}^{\text{TB}} \rangle\,,
\end{split}
\label{eq:downfolded_H_and_S}
\end{equation}
and a corresponding orthogonal TB Hamiltonian is obtained by L\"owdin transformation~\cite{doi:10.1080/00018735600101155}
\begin{equation}
H^{\text{TB}, \text{orth}} = S^{-1/2} H^{\text{TB}} S^{-1/2} \,.
\label{eq:loewdin}
\end{equation}

\subsection{Parameterization\label{sec:parameterization}}

The asymptotic behavior of the distance dependence of the Hamiltonian matrix elements is well described by an exponential decay. A function to parameterize the bond integrals should capture the asymptotic behavior and at the same time be sufficiently flexible to model the matrix elements at shorter interatomic distance. We choose to parameterize the bond integrals as a sum of exponentials,
\begin{equation}
\beta(R) = \sum_{i=0}^{i_{\text{max}}} c_{i} \exp \left(-\lambda_i R^{n_{i}}\right) = \sum_{i=0}^{i_{\text{max}}} f_i(R)
\label{eq:beta}
\end{equation}
with $n_0=1$. The diagonal onsite matrix elements are parameterized by the same functional form,
\begin{equation}
E(R) = \sum_{i=0}^{i_{\text{max}}} c_{i} \exp \left(-\lambda_i R^{n_{i}}\right) = \sum_{i=0}^{i_{\text{max}}} f_i(R)
\label{eq:onsite}
\end{equation}
with $\lambda_0 = 0$, which sets the first term to a constant value.

The parameterization was carried out using the following procedure :
\begin{itemize}
\item Define a threshold $\Delta$ which is equal to the largest allowed quadratic difference between the fit and the raw data.
\item Find the smallest interatomic distance up to which $f_0(R)$ can describe the raw data without exceeding the threshold $\Delta$.
\item Subtract $f_0(R)$ from the raw data and fit the remaining data with $f_1(R)$ up to the smallest interatomic distance for which the threshold $\Delta$ is exceeded.
\item Continue by increasing $i$ to $i_{\text{max}}$ until the fit $\sum_i f_i(R)$ can accurately describe all data points up to $\min \left( 3/4 R_{\text{eq}}, 3 \text{\AA} \right)$. 
\end{itemize}

\section{\label{sec:Si_analysis}Si-Si and Mo-Mo as examples of $sp$-$sp$ and $sd$-$sd$ valent dimers}

We show results for Si and Mo as representative examples of $sp$-valent and $sd$-valent dimers. Matrix elements for all dimers are available in the Supplemental Material at Ref.~\cite{SuppMat}. 
\begin{figure}[htb]
\includegraphics[width=0.9\columnwidth]{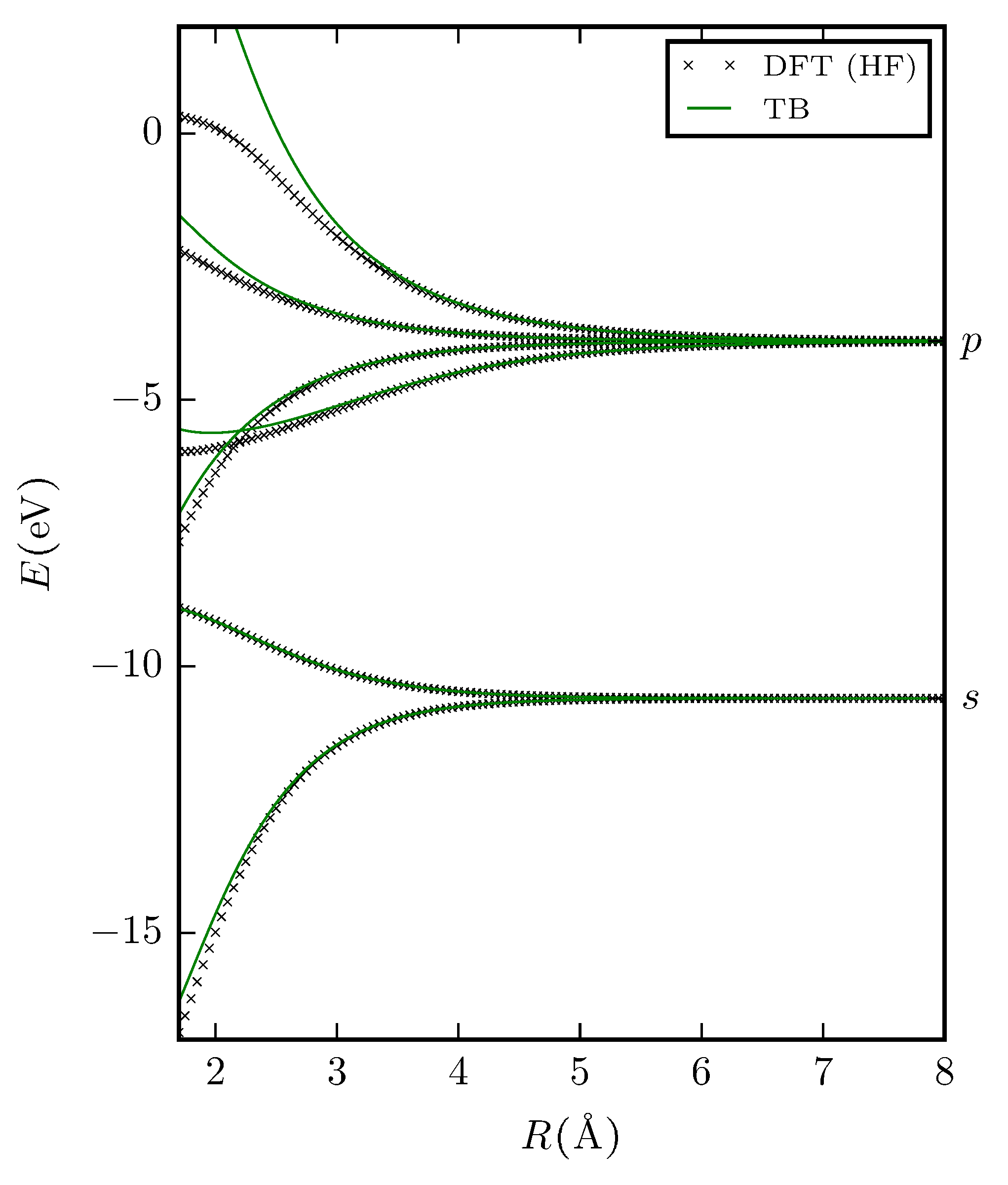}\\
\includegraphics[width=0.9\columnwidth]{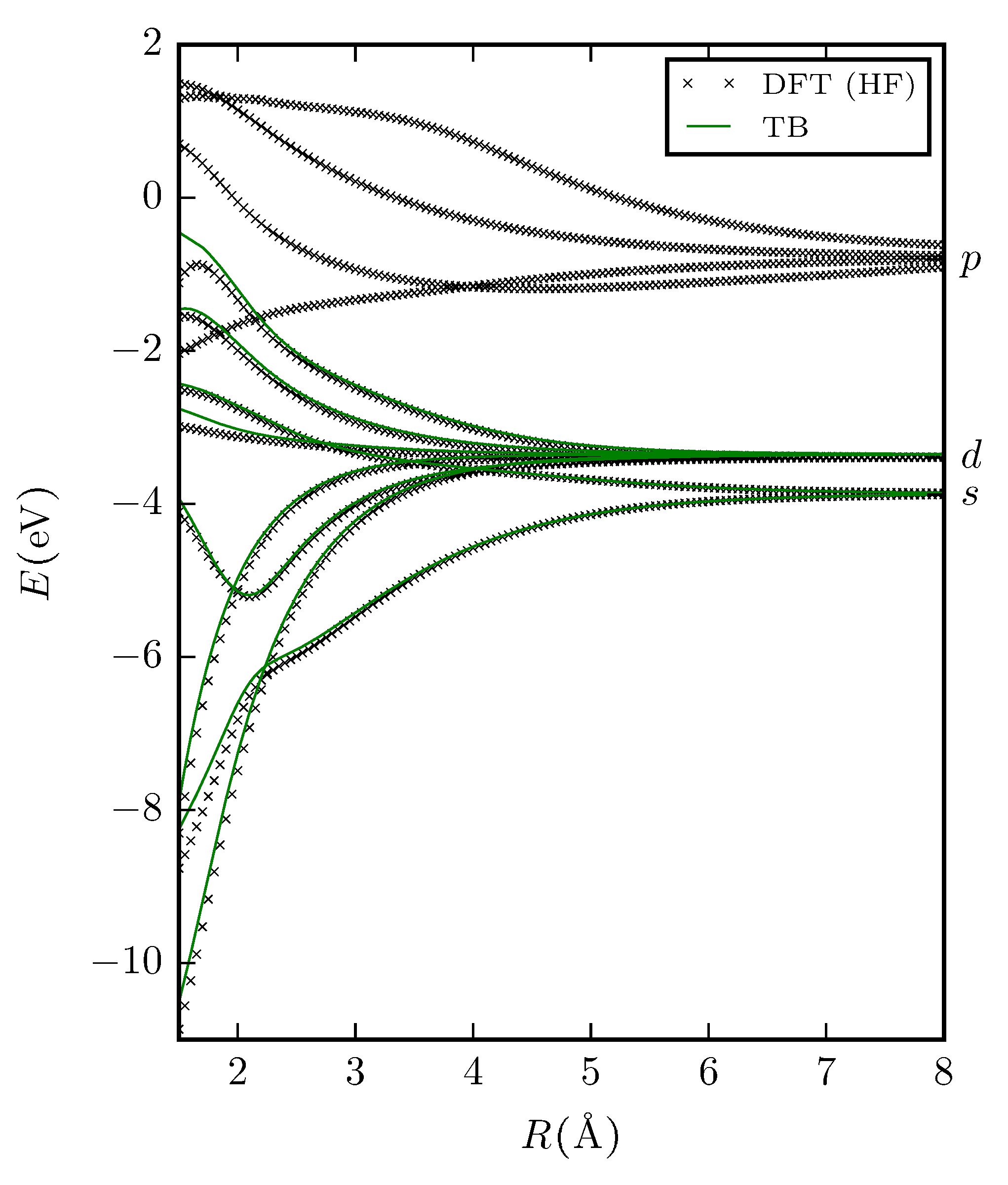}
\caption{TB and DFT eigenvalues as a function of distance for the Si dimer (top) and the Mo dimer (bottom). } 
\label{fig:SiMo_spectrum}
\end{figure}
The distance dependence of the eigenvalues of the Si and Mo dimers as obtained from HF-DFT and the TB bond parameters are shown in Fig.~\ref{fig:SiMo_spectrum}. A $sp$ and $sd$ valence was used for Si and Mo, respectively.
Note that the TB eigenspectrum in Fig.~\ref{fig:SiMo_spectrum} was computed with the downfolded Hamiltonian includingtwo-center onsite levels that are omitted in the following parameterization.

For Si, two $\sigma$-states that are formed predominately from combinations of the $s$-orbitals are lower in energy than the other two $\sigma$ and four $\pi$-states that are formed largely from $p$-orbitals. 
The states of the $\pi$-block are two-fold degenerated. 
The TB eigenenergies are in good agreement with HF-DFT for large interatomic distances where the atomic orbitals are similar to those of free atoms. 
For shorter interatomic distances the agreement for the occupied states is also good while the TB eigenenergies of the unoccupied states show deviations from HF-DFT. 
The latter is a consequence of weighting the optimal projection with the occupation number (Eq.~\ref{eq:max_P}):
The minimal basis, which cannot reproduce all states exactly, is chosen such that the occupied states are well reproduced. 

The eigenvalues obtained with the TB bond parameters for the Mo dimer reproduces the HF-DFT reference very well, also. Compared to Si, the Mo dimer has four additional two-fold degenerated $\delta$-states eigenvalues, and the HF-DFT $p$-valence-states are also shown in Fig.~\ref{fig:SiMo_spectrum}. 

\begin{figure*}[htb!]
\includegraphics[width=0.9\textwidth]{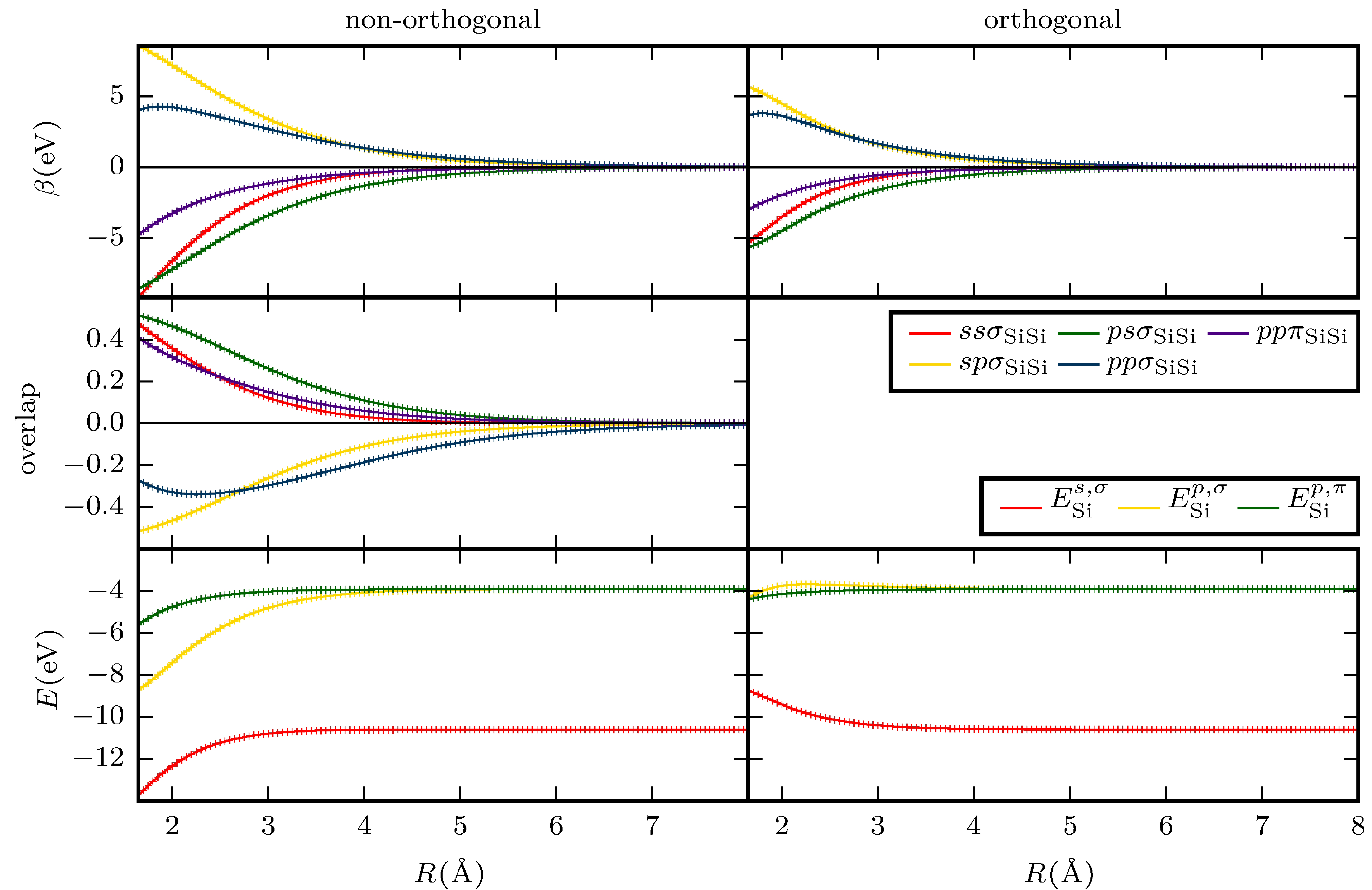}
\vspace{0.5cm}\\
\includegraphics[width=0.9\textwidth]{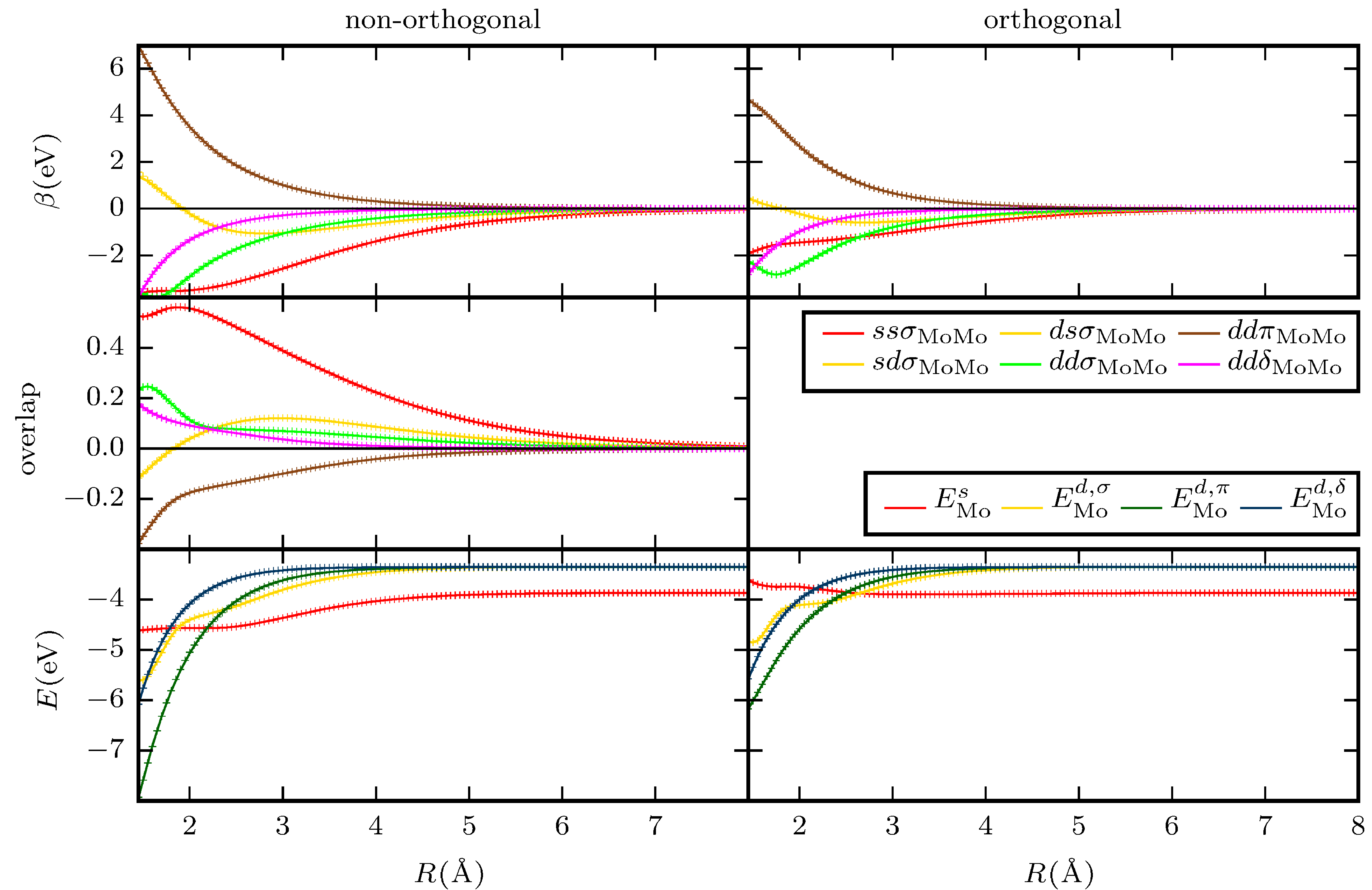}
\caption{Matrix elements of non-orthogonal TB Hamiltonian matrix, overlap matrix and onsite levels obtained by downfolding and orthogonal TB Hamiltonian matrix for Si (top) and Mo (bottom) dimer. 
Symbols correspond to the matrix elements from downfolding, solid lines are interpolations using Eqs.~\ref{eq:beta} and~\ref{eq:onsite}.}
\label{fig:SiMo}
\end{figure*}

The distance-dependence of the individual elements of the TB Hamiltonian matrix, the overlap matrix and the L\"owdin-orthogonalized TB Hamiltonian matrix are shown in Fig.~\ref{fig:SiMo}. 
All matrix elements of the Si-Si dimer are parametrized with three exponential terms, i.e. $i_{\text{max}} = 2$ in Eqs.~\ref{eq:beta} and~\ref{eq:onsite}, while six terms are used for the Mo dimer.
This is a consequence of the comparably more complex eigenspectrum of Mo-Mo at small interatomic distances (Fig.~\ref{fig:SiMo_spectrum}).
The excellent agreement of parameterized and downfolded TB matrix elements in Fig.~\ref{fig:SiMo} shows that the errors from the parameterization procedure with sums of exponentials (Eqs.~\ref{eq:beta} and~\ref{eq:onsite}) can be neglected.

\section{\label{sec:homo}Trend across homoatomic dimers}

\begin{figure*}[htb]
\includegraphics[width=0.9\columnwidth]{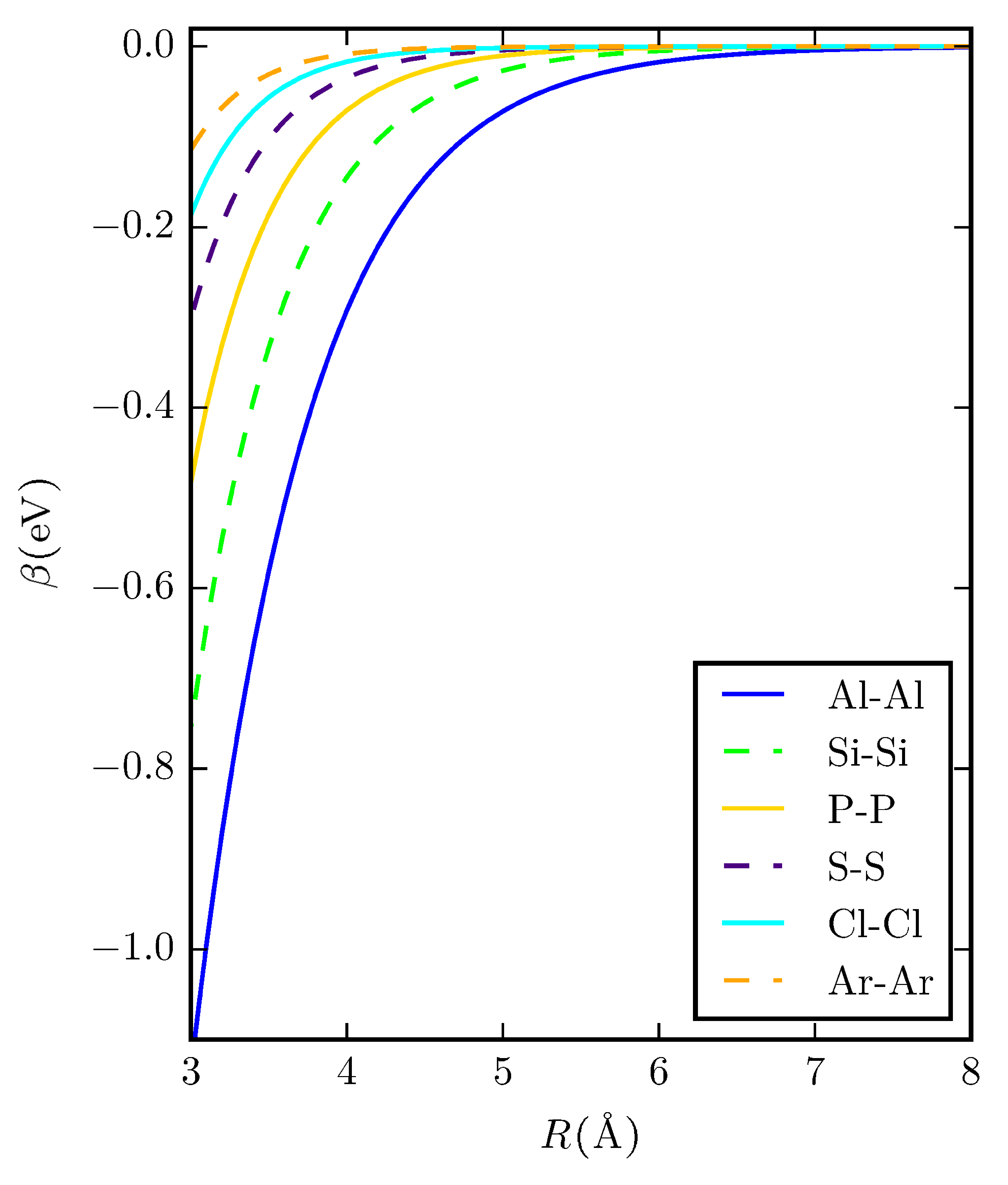}
\qquad
\includegraphics[width=0.9\columnwidth]{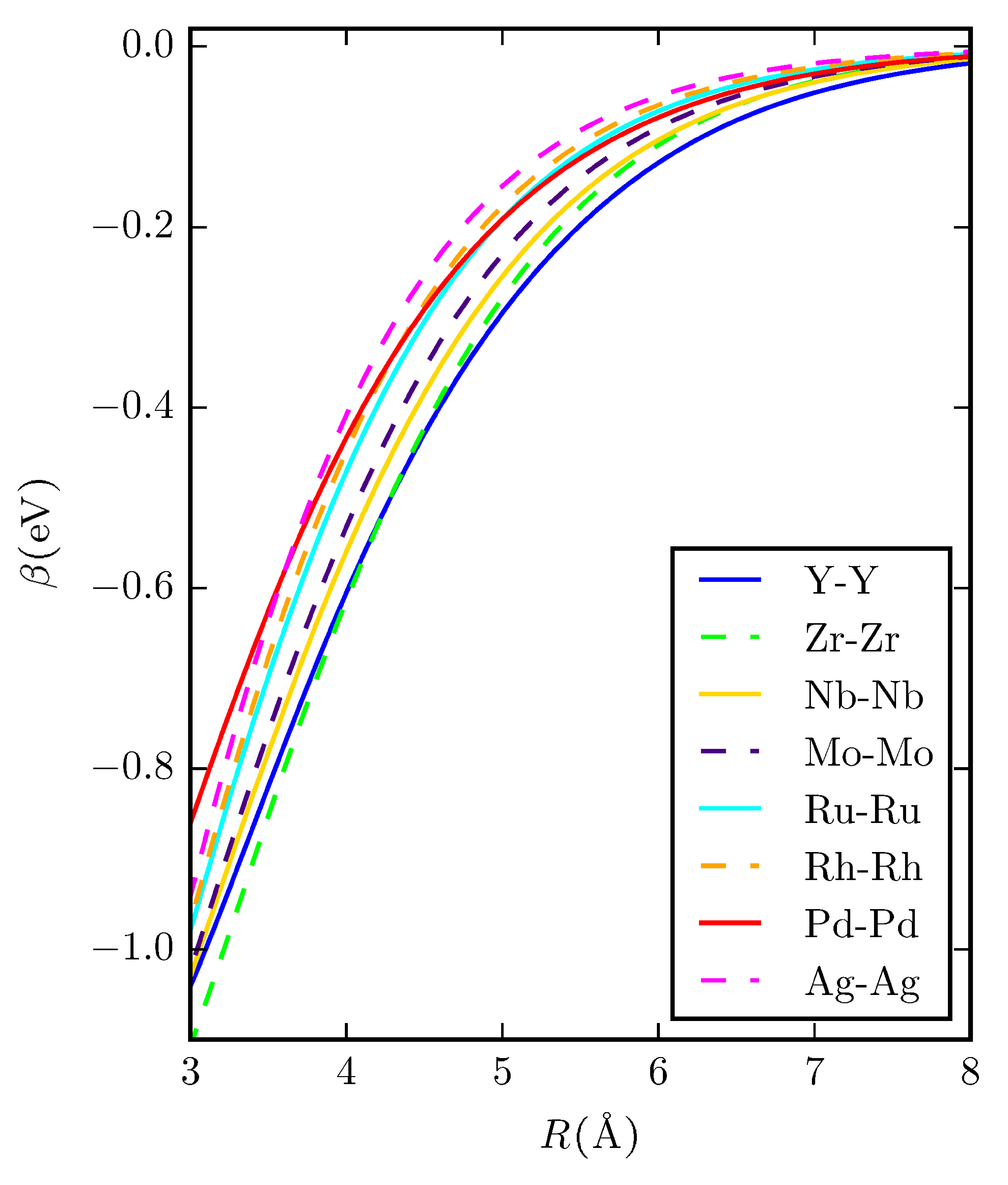}
\caption{Bond integrals $ss\sigma$ of orthogonal TB models across period $3p$ (left) and period $4d$ (right).}
\label{fig:sss_H_orth}
\end{figure*}

\begin{figure*}[htb]
\includegraphics[width=0.9\columnwidth]{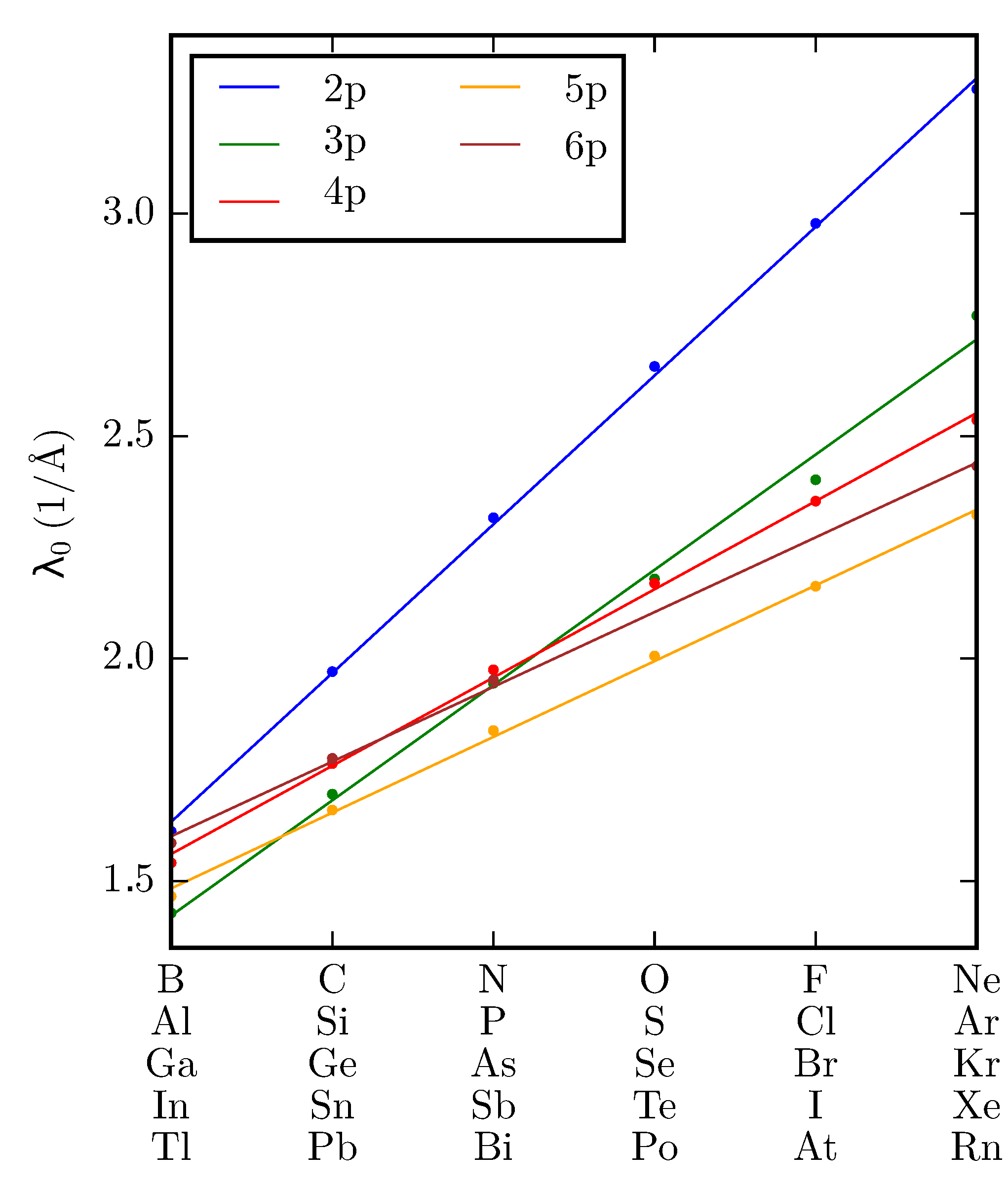}
\qquad
\includegraphics[width=0.9\columnwidth]{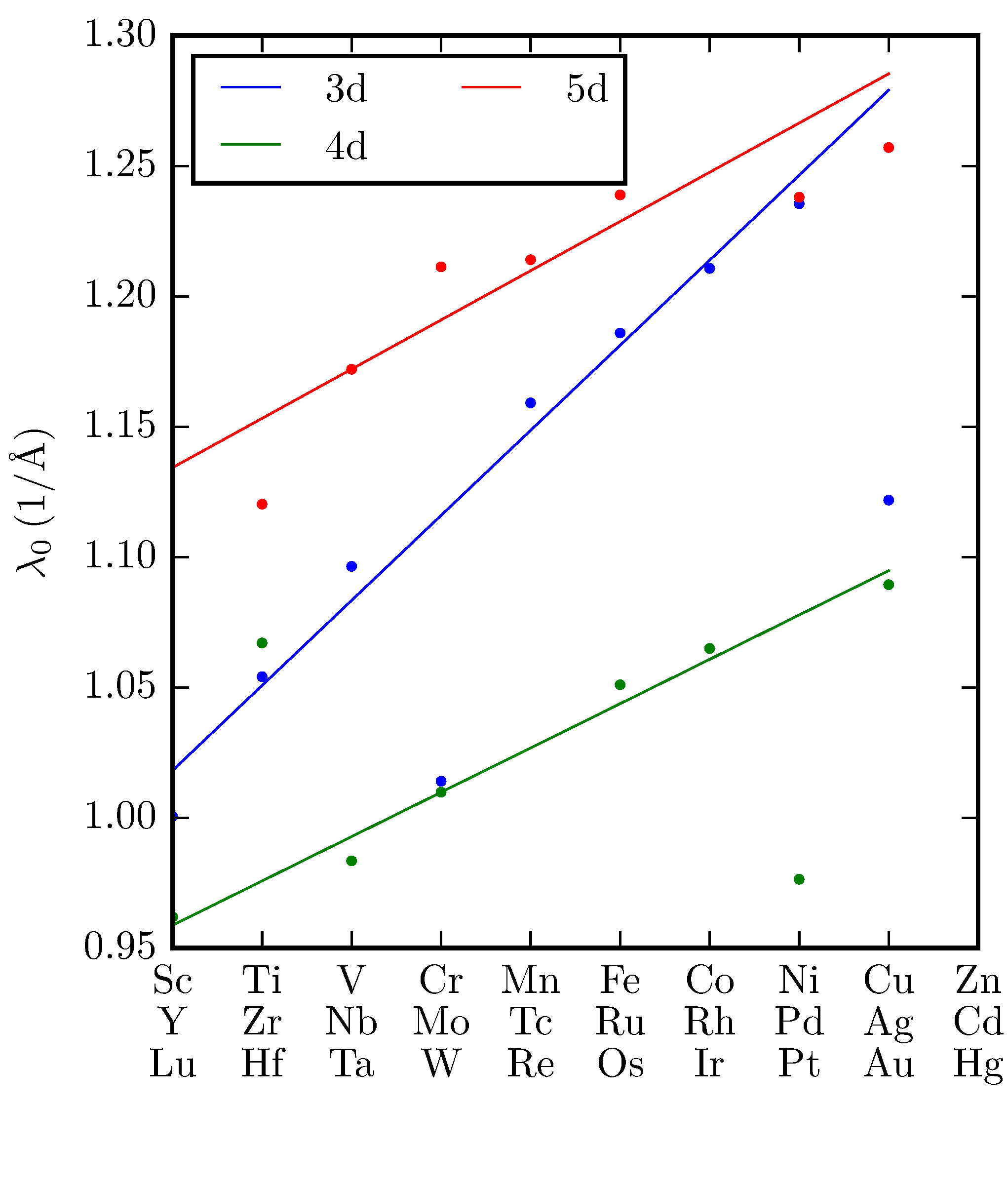}
\caption{Inverse decay length $\lambda_0$ of the $ss\sigma$ matrix elements of orthogonal TB models across the periods of the $sp$-valent (left) and $sd$-valent (right) homoatomic dimers.}
\label{fig:p1_H_orth_homo}
\end{figure*}

In Fig.~\ref{fig:sss_H_orth} we show the $ss\sigma$ matrix element of the orthogonalized Hamiltonian for all homoatomic dimers of period $3p$ and $4d$.  
The magnitude of the bond integrals from Al to Ar decreases by an order of magnitude. This is a consequence of the increasing nuclear charge across the period which leads to a contraction of core and valence orbitals. The homoatomic dimers of period $4d$ show a similar overall behavior with a much smaller variation across the period while the range of the bond integrals is longer. A numerical measure of the range of the bond integrals is given by the coefficient $\lambda_0$ of our parametrization function (Eq.~\ref{eq:beta}) that corresponds to an inverse decay length~\cite{PETTIFOR198743}.
The values of $\lambda_0$ across the different periods are shown in Fig.~\ref{fig:p1_H_orth_homo} for the $ss\sigma$ bond integral of the orthogonalized Hamiltonian.
The inverse decay length of the homoatomic bond integrals across the different periods is well described by a linear relationship,
\begin{equation}
\lambda_0 = b_0 + m_0 N_{p/d},
\label{eq:lin_reg}
\end{equation}
where $N_{p/d}$ is the number of $p/d$ electrons of the atoms.
The values of $m_0$ and $b_0$ as obtained by linear regression are summarized in Tab.~\ref{tab:lin_reg}.
The larger slope $m_0$ of $ss\sigma$ for the $sp$-elements means a faster decay as compared to the $sd$-valent elements.

\begin{table*}[ht]
\caption{
Slope $m_0$ and intercept $b_0$ of the inverse decay length $\lambda_0$ of  bond integrals of orthogonal TB models across the periods.}
\begin{ruledtabular}
\begin{tabular}{c|ccccc|ccccc}
&\multicolumn{5}{c}{$m_0/(1/\text{\AA})$}&\multicolumn{5}{c}{$b_0/(1/\text{\AA})$}\\

period & $ss\sigma$ & $sp\sigma$ & $pp\sigma$ & $pp\pi$ & & $ss\sigma$ & $sp\sigma$ & $pp\sigma$ & $pp\pi$ & \\
\hline
$2p$ & 0.334 & 0.261 & 0.207 & 0.222 & & 1.299 & 0.859 & 0.843 & 1.032 & \\
$3p$ & 0.259 & 0.221 & 0.179 & 0.191 & & 1.164 & 0.697 & 0.700 & 0.864 & \\
$4p$ & 0.198 & 0.196 & 0.163 & 0.170 & & 1.363 & 0.740 & 0.696 & 0.868 & \\
$5p$ & 0.170 & 0.146 & 0.145 & 0.148 & & 1.313 & 0.801 & 0.689 & 0.866 & \\
$6p$ & 0.168 & 0.136 & 0.138 & 0.139 & & 1.433 & 0.781 & 0.668 & 0.847 & \\
\hline
     & $ss\sigma$ & $sd\sigma$ & $dd\sigma$ & $dd\pi$ & $dd\delta$ & $ss\sigma$ & $sd\sigma$ & $dd\sigma$ & $dd\pi$ & $dd\delta$\\
\hline
$3d$ & 0.033 & 0.057 & 0.112 & 0.107 & 0.123 & 0.985 & 0.927 & 0.882 & 1.094 & 1.257 \\
$4d$ & 0.017 & 0.041 & 0.101 & 0.099 & 0.105 & 0.942 & 0.854 & 0.684 & 0.877 & 1.044 \\
$5d$ & 0.019 & 0.049 & 0.087 & 0.090 & 0.092 & 1.115 & 0.961 & 0.823 & 0.997 & 1.176 \\
\end{tabular}
\end{ruledtabular}
\label{tab:lin_reg}
\end{table*}

The non-monotonic variation of $m_0$ with the period of the periodic table can be attributed to the change of the size of the atomic core with the electronic configuration.
The fact that the values of $m_0$ are close for $dd\sigma$, $dd\pi$ and $dd\delta$ may be viewed as a justification for using the same inverse decay length for the $dd\sigma$, $dd\pi$ and $dd\delta$ bond integrals in the rectangular $d$-band model of Pettifor~\cite{PETTIFOR198743}. The numerical values for  $m_0$  and $b_0$ in the rectangular $d$-band model ($m_{0, d\text{-band}} = 0.142/\text{\AA}$, $b_{0, d\text{-band}} = 0.966/\text{\AA}$) are slightly larger than our results for the $4d$-period. This may reflect the faster decay of the two-center bond integrals in the bulk crystal due to screening by neighboring atoms~\cite{PhysRevLett.85.4136, DRAUTZ2007196, 0953-8984-25-11-115502, PhysRevB.89.235134}.
\begin{figure*}[htb]
\includegraphics[width=0.7\textwidth]{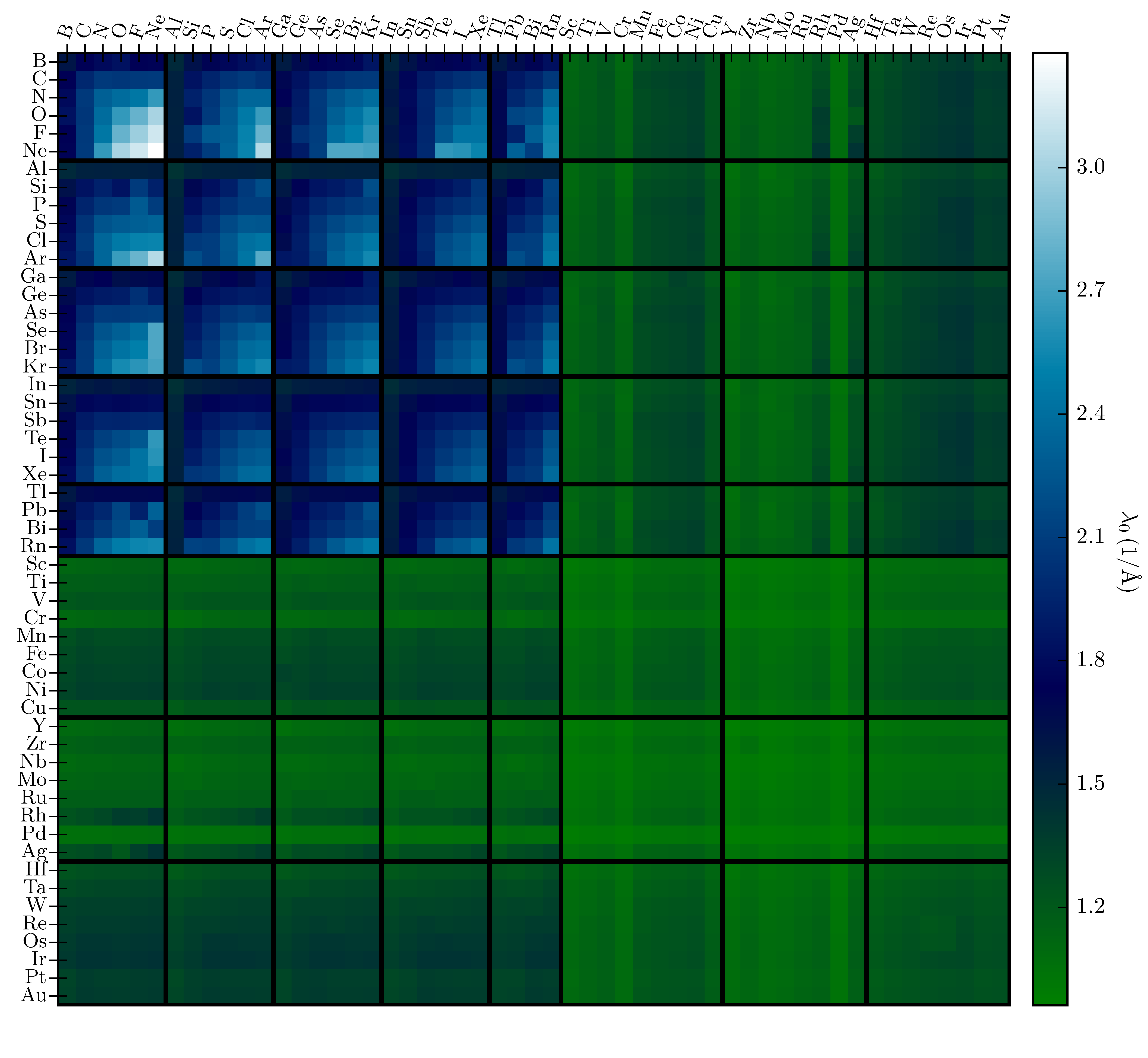}
\caption{Inverse decay length $\lambda_0$ for $ss\sigma$ and orthogonal TB models across homoatomic and heteroatomic dimers.}
\label{fig:p1_H_orth_all_elements}
\end{figure*} 

The elements Cr, Cu, Zr and Pd were excluded from the linear regression (Eq.~\ref{eq:lin_reg}) as their values of $\lambda_0$ and of all TB matrix elements are clear outliers among the $sd$-valent homoatomic dimers. Similar deviations are observed for heteroatomic dimers that include these elements, see Sec.~\ref{sec:hetero}. We attribute the deviations to the respective GPAW datasets. Their asymptotic behavior from the dimer to the free atom is apparently different for Cr, Cu, Zr and Pd as the respective inverse decay lengths $\lambda_0$ (Eq.~\ref{eq:beta}) deviate from the overall trend. These outliers in the limit of the neutral free-atom lead to deviations in the input density for the HF approximation which leads to the outliers in the TB parameters.

\section{Trend across heteroatomic dimers\label{sec:hetero}}

Figure~\ref{fig:p1_H_orth_all_elements} shows the inverse decay length $\lambda_0$ of the $ss\sigma$ bond integral of orthogonal TB models for heteroatomic dimers of period 2 to 6. 
The value of $\lambda_0$ is determined by both, the period and the number of valence electrons of the two atoms. 
(The elements Cr, Cu, Zr or Pd that appeared as outliers in homoatomic dimers are also outliers in the trends across heteroatomic dimers.)
As expected from the results for the homoatomic dimers (Tab.~\ref{tab:lin_reg}), we find that the $sp$-$sp$ dimers exhibit the largest values of $\lambda_0$, i.e. the shortest-ranged orbitals, and the largest variation across the period. 
The $ss\sigma$ bond integrals of the heterovalent $sp$-$sd$ dimers are longer ranged and similar to the $sd$-$sd$ dimers. 
The variation of $\lambda_0$ for $sp$-$sp$ and $sd$-$sd$ dimers of elements of different periods is driven by the number of valence electrons of each atom.
The change of $\lambda_0$ for $sp$-$sd$ dimers, in contrast, is determined mostly by the number of valence electrons of the $sd$-element with little importance of the $sp$-element. 
The small variation of $\lambda_0$ for $sd$-$sd$ dimers indicates that a TB model for $sd$-$sd$ compounds could in good approximation assume a constant decay length of the orbitals across the period.

\section{Comparison to available TB models \label{sec:compare}}

In the following we compare the parameterizations obtained in this work to available TB parameterizations. Simple TB models have been used for many years to rationalize observations from experiment or electronic structure calculations. We compare the results from our downfolding procedure to the reduced TB approximation and to canonical TB models for $sp$- and $d$-valent materials. We further give a brief comparison to other TB methods, namely NRL-TB and DFTB. The parameterization of TB bond parameters in this work covers only the contribution of the bond energy to the TB total energy. Therefore the following discussion is based on comparisons of the TB bond parameters and the resulting electronic density of states (DOS). For a detailed discussion of the physical origin and the parameterization of all contributions we refer the reader to recent descriptions of TB/BOP parameterizations~\cite{0965-0393-23-7-074004,Ferrari-19,Ladines-20}.

\subsection{Reduced TB approximation}

\begin{figure*}[htb]
\includegraphics[width=0.9\columnwidth]{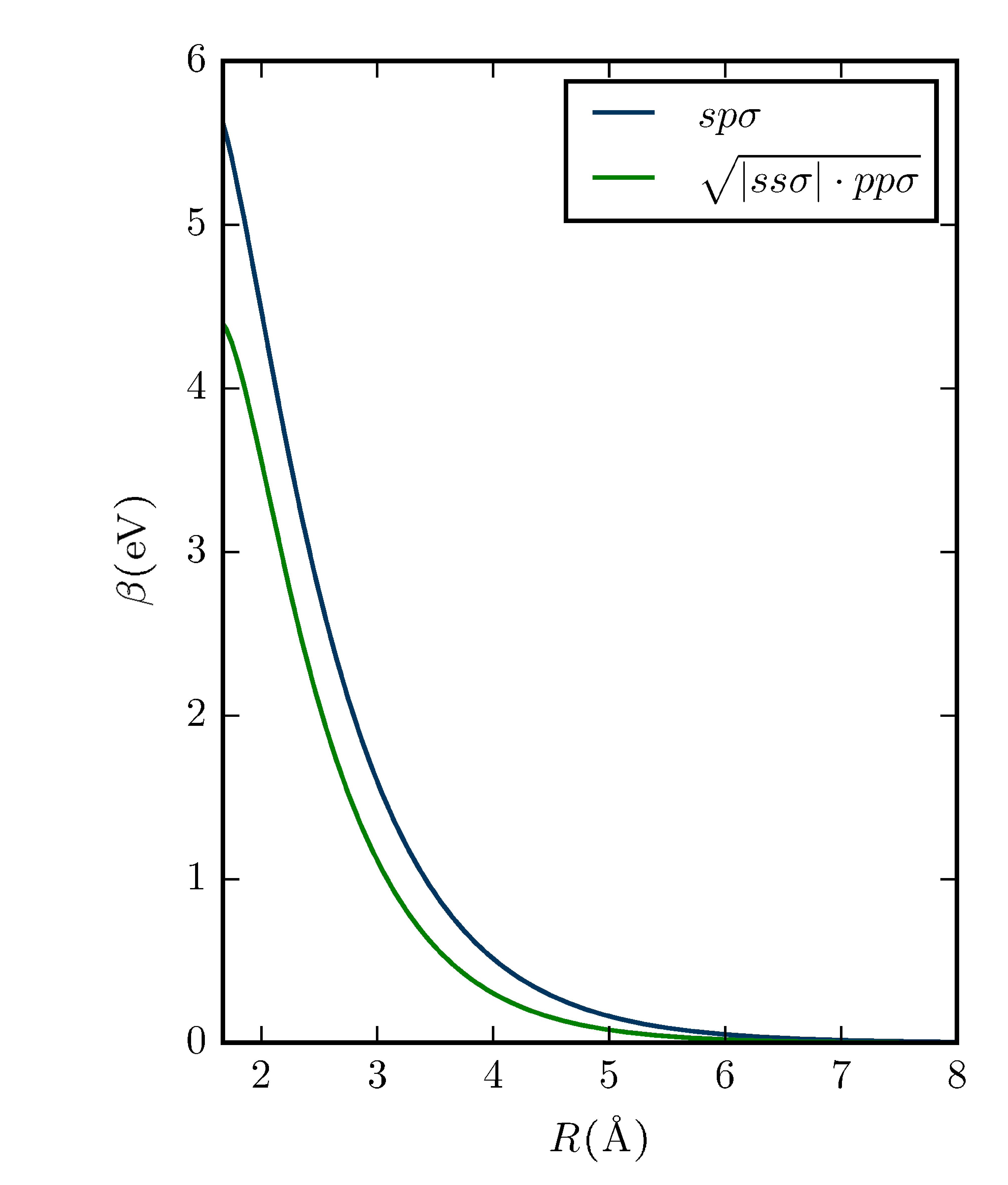}
\qquad
\includegraphics[width=0.9\columnwidth]{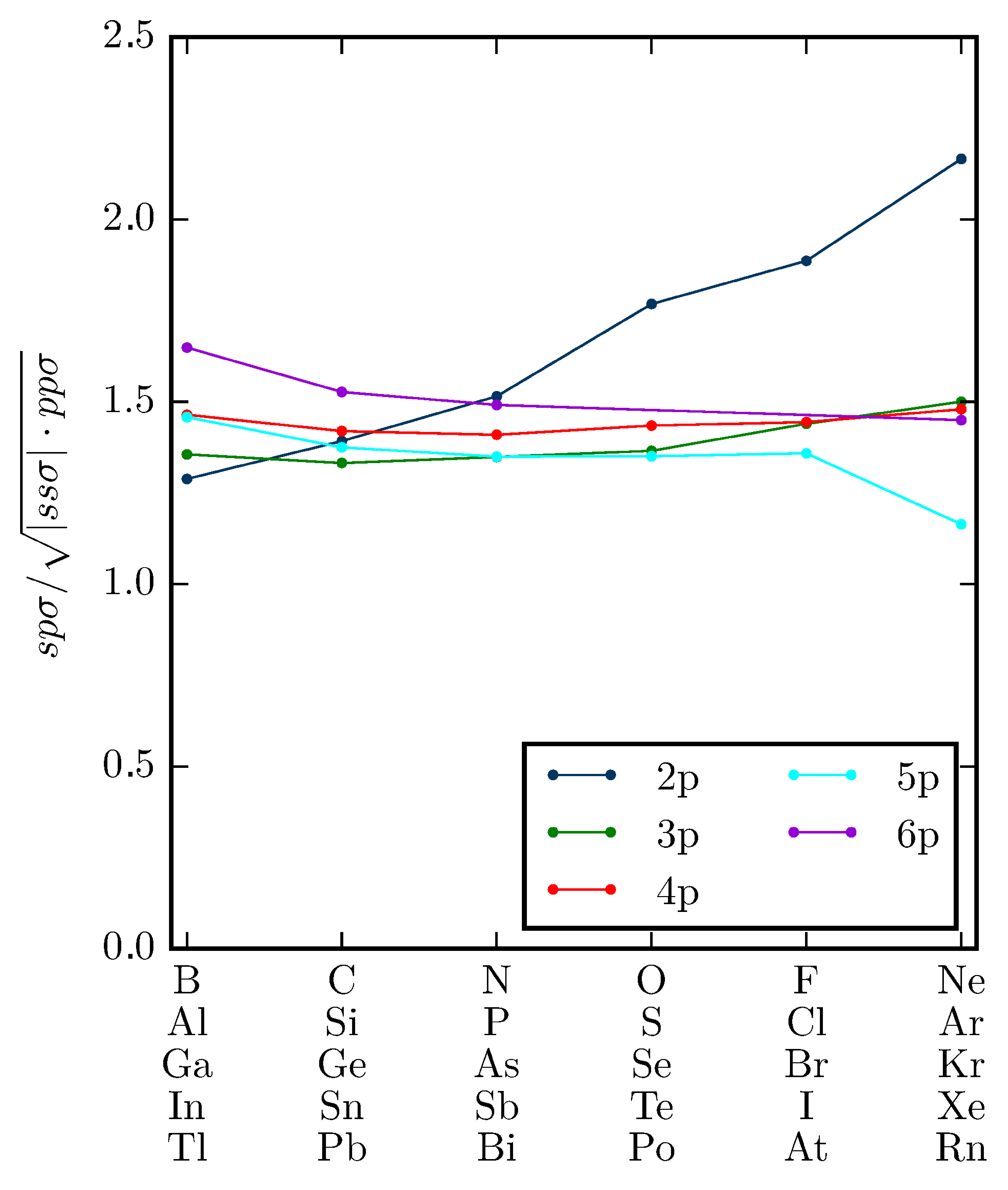}
\caption{Bond integral $sp\sigma$ and reduced TB approximation for Si (left) and across the $sp$-block (right) at a fixed interatomic distance of $R=2.5\mathrm{\AA}$.} 
\label{fig:reduced_TB_approx}
\end{figure*}

The reduced TB approximation for $sp$-valent elements~\cite{PhysRevB.59.8487, PhysRevB.59.8500, PhysRevB.91.054109} approximates the $sp\sigma$ bond integral as geometric mean
\begin{equation}
sp\sigma = \sqrt{|ss\sigma| \cdot pp\sigma} \,,
\label{eq:reducedTB} 
\end{equation}
of $ss\sigma$ and $pp\sigma$. 
In Fig.~\ref{fig:reduced_TB_approx} the $sp\sigma$ bond integral is compared to $\sqrt{|ss\sigma| \cdot pp\sigma}$ for the the Si dimer.
The reduced TB approximation is in qualitative agreement regarding the overall distance-dependence but underestimates the value of the $sp\sigma$ bond integral.
As shown in the right hand panel of Fig.~\ref{fig:reduced_TB_approx}, this observation generalizes to $sp$-valent elements. The dimers are compared at a length of $R=2.5\mathrm{\AA}$. The ratio $sp\sigma / \sqrt{|ss\sigma| \cdot pp\sigma}$ across the $sp$-block varies only slightly, except for elements of period 2$p$ that  do not have $p$ core states. Overall a value of $sp\sigma \approx 1.5 \sqrt{|ss\sigma| \cdot pp\sigma}$ would provide a better quantitative agreement with the downfolded $sp\sigma$ bond integrals.

\subsection{Canonical TB models}

\begin{figure*}[htb!]
\includegraphics[width=0.9\columnwidth]{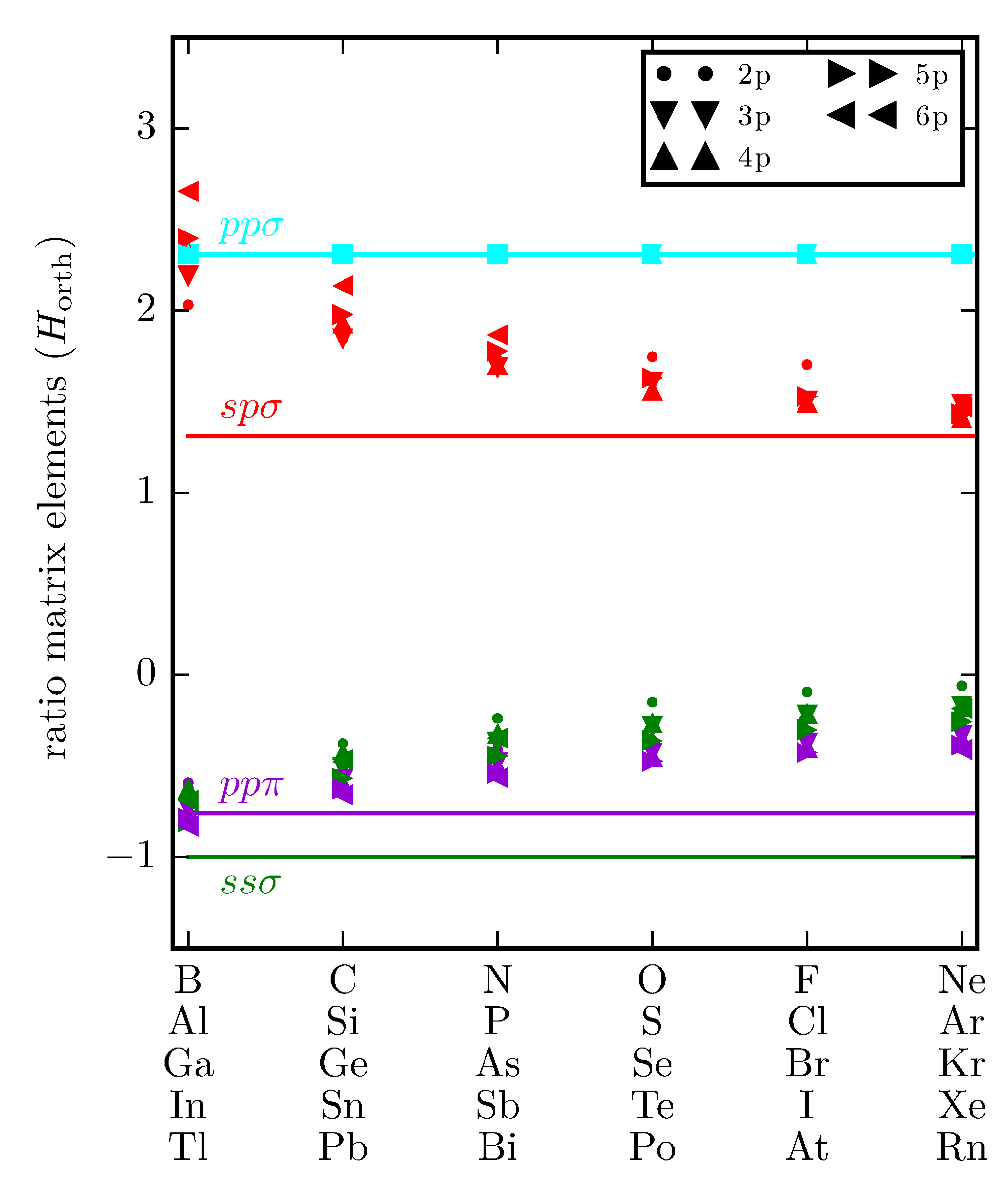}
\label{fig:ratio_bondintegrals_sp}
\includegraphics[width=0.9\columnwidth]{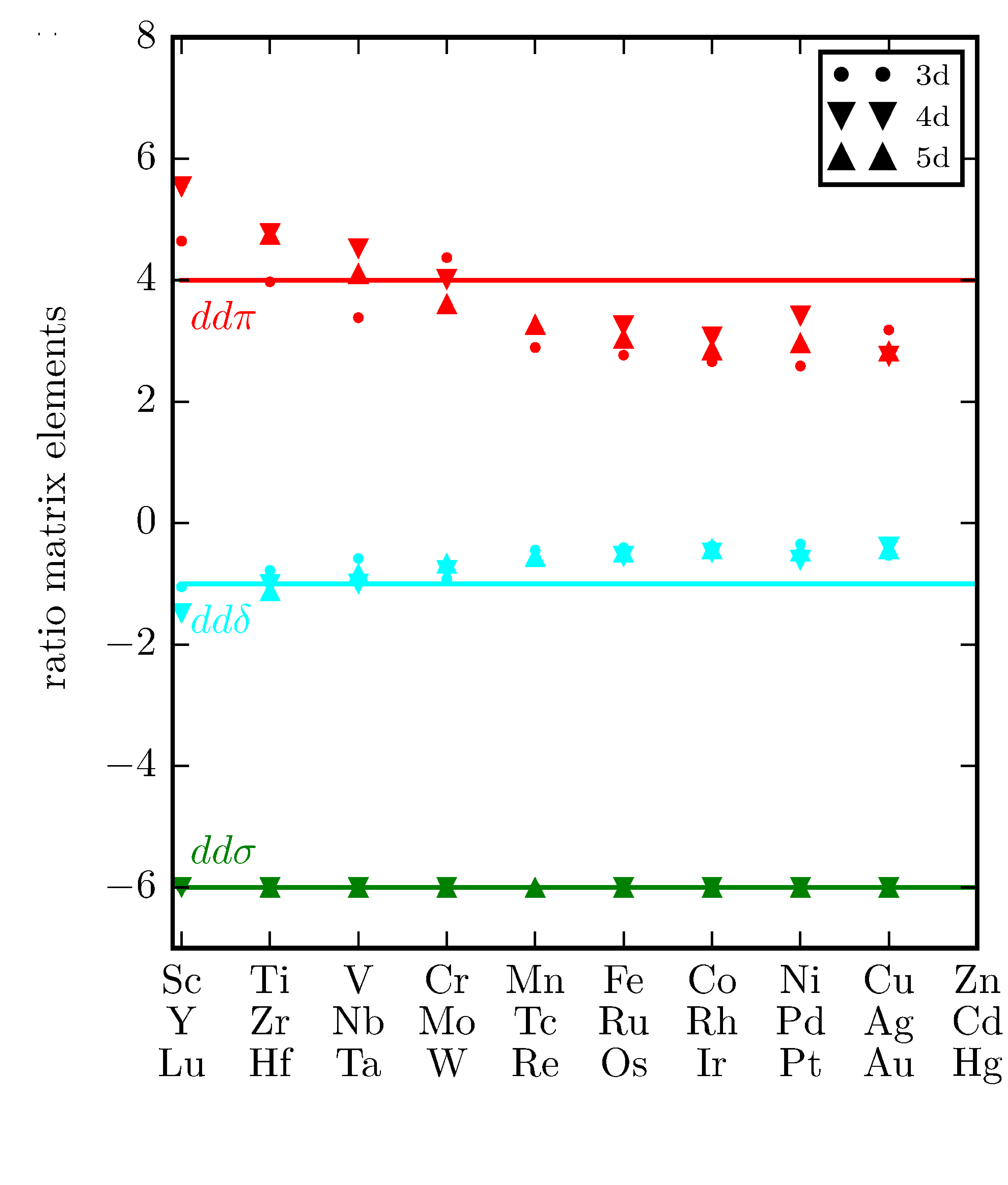}
\label{fig:ratio_bondintegrals_d}
\caption{Bond integrals of homoatomic dimers compared to canonical TB models, $sp$-valent (left) and $d$-valent (right). Horizontal lines correspond to the canonical TB models. (right panel: solid lines: Eq.~\ref{eq:TB_Anderson}, dashed lines: Eq.~\ref{eq:TB_Turchi}).}
\label{fig:ratio_bondintegral}
\end{figure*}

Canonical TB models assume a constant relative ratio of the bond integrals in $sp$- or $d$- valent materials~\cite{0953-8984-3-5-001,Turchi-91,PhysRevB.17.1209}.
The canonical $sp$ model of Cressoni and Pettifor~\cite{0953-8984-3-5-001} captures the structural trends across the $sp$-valent materials by assuming
\begin{equation}
pp\pi : ss\sigma : sp\sigma : pp\sigma  = -0.76:-1:1.31:2.31 \,.
\label{eq:TB_Cressoni}
\end{equation}
Further, the radial functions across the elements are taken to have the same radial decay, an approximation that does not agree with the systematic change of the inverse decay length across the elements discussed in Sec.~\ref{sec:homo}. 
In the following we therefore determine the ratios of the bond integrals at a fixed bond length of $R=4\mathrm{\AA}$. At this distance the decay of the matrix elements is well described by an exponential decay. The ratios of the bond integrals in an orthogonal TB model are shown in Fig.~\ref{fig:ratio_bondintegral}. We divide the bond integrals by $pp\sigma$ and multiply with the corresponding value of $pp\sigma=2.31$ in the canonical model. 

The ordering of the downfolded TB bond parameters is different from the canonical TB model.  The relative ordering of the matrix elements is given by
\begin{equation}
|ss\sigma|<|pp\pi|<|sp\sigma|<|pp\sigma| \,,
\end{equation}
for all elements of the $sp$ block except for In and Tl where $|sp\sigma|$ is largest.
For early $sp$ elements, the ratio of $pp\pi$ and $ss\sigma$ to $pp\sigma$ are in good agreement with the canonical $sp$ model, while the ratio of $sp\sigma : pp\sigma$ is considerably off. 
This changes for late $sp$ elements, where the canonical TB model agrees for $sp\sigma$ but not  $pp\pi$ and $ss\sigma$. 
We note that applying the reduced TB approximation (Eq.~\ref{eq:reducedTB}) to the canonical ratio of $ss\sigma$ and $pp\sigma$ leads to $sp\sigma/ss\sigma=1.52$, which improves the overall agreement with the downfolded dimer matrix elements. 

Simple canonical TB models for transition metals were shown to provide good structural energy differences for intermetallics across the transition metal series, see, e.g., ~\cite{Hammerschmidt-08-2,PhysRevB.83.224116}. For transition metals two flavors of canonical TB models are used. Andersen {\it et al.}~\cite{PhysRevB.17.1209} assume
\begin{equation}
dd\sigma : dd\pi : dd\delta = -6:4:-1
\label{eq:TB_Anderson}
\end{equation}
while Turchi~\cite{Turchi-91} uses
\begin{equation}
dd\sigma : dd\pi : dd\delta = -2:1:0.
\label{eq:TB_Turchi}
\end{equation}
Both canonical TB models are compared to the numerical ratios of the $dd$ matrix elements of orthogonal TB models in Fig.~\ref{fig:ratio_bondintegral}, where the bond integrals were scaled to match $dd\sigma=6$. 
Despite considerable variations in the downfolded TB matrix elements, we observe a good overall agreement with the two $d$-valent canonical TB models. The ratios of the downfolded bond integrals are close to the canonical TB model of Andersen {\it et al.}~\cite{PhysRevB.17.1209} particularly for the first half of the series and to the canonical model of Turchi~\cite{Turchi-91}  for the second half of the series.

As for the $sp$-valent dimers, we find very similar ratios of the bond integrals for the same group of the different series. The variation among the same group is considerably smaller than both, the ratio of the bond-integrals and their variation across the series. The elements Cu, Cr, Pd and Zr appeared as outliers with regard to the values of the matrix elements in Sec.~\ref{sec:homo} while in Fig.~\ref{fig:ratio_bondintegral} they are fully in line with the trend of the ratios of the matrix elements. We interpret this as further indication that the asymptotic limit of the GPAW datasets for these elements is somewhat inconsistent to the other elements while the relative contribution of the different orbitals to the interatomic bond matches the trend.

\subsection{NRL-TB and DFTB}

We compare the TB parameters obtained in this work to NRL-TB~\cite{0953-8984-15-10-201} and DFTB~\cite{PhysRevB.51.12947, PhysRevB.58.7260}.
In the NRL-TB formalism, the repulsive interaction between atoms is modeled by a shift of the one-electron eigenvalues. Parametrization of the bond integrals for all homoatomic systems of the $d$-block~\cite{PhysRevB.54.4519} as well as a parametrization of ground-state structures across the periodic table~\cite{papaconstantopoulos1986handbook} are available. 
These parameterizations were obtained by direct fitting of the Hamiltonian matrix elements to DFT reference data, which includes total energies and band structures~\cite{PhysRevB.74.054104}. 
In DFTB, pseudoatomic wave functions are defined using a confinement potential. 
The Hamiltonian matrix elements are computed in the two-center approximation from pseudoatomic wave functions. This corresponds to calculating the TB matrix elements from the pseudoatomic wave functions for a dimer Hamiltonian. 
The parameters of the confinement potential are optimized to reproduce selected reference data. 
A DFTB parametrization across the periodic table has been obtained by fitting the model parameters to unary bulk structures by Wahiduzzaman {\it et al.}~\cite{doi:10.1021/ct4004959} and the performance of the model parameters was tested for binary systems. Grimme {\it et al.} parametrized the GFN-xTB Hamiltonian across the periodic table~\cite{doi:10.1021/acs.jctc.7b00118} in a similar way. 

\begin{figure*}[htb!]
\includegraphics[width=0.9\columnwidth]{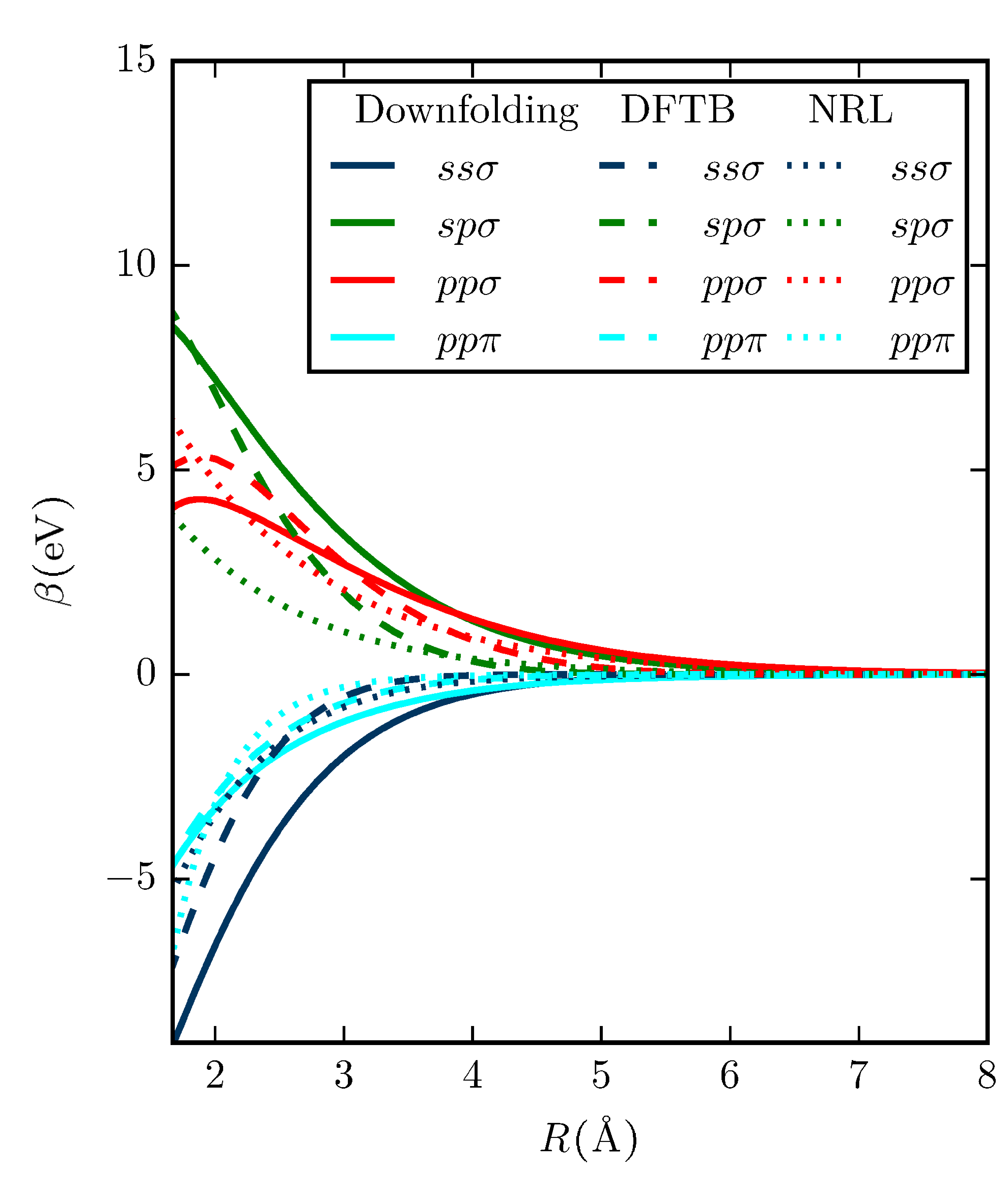}
\label{fig:comp_TB_H}
\qquad
\includegraphics[width=0.9\columnwidth]{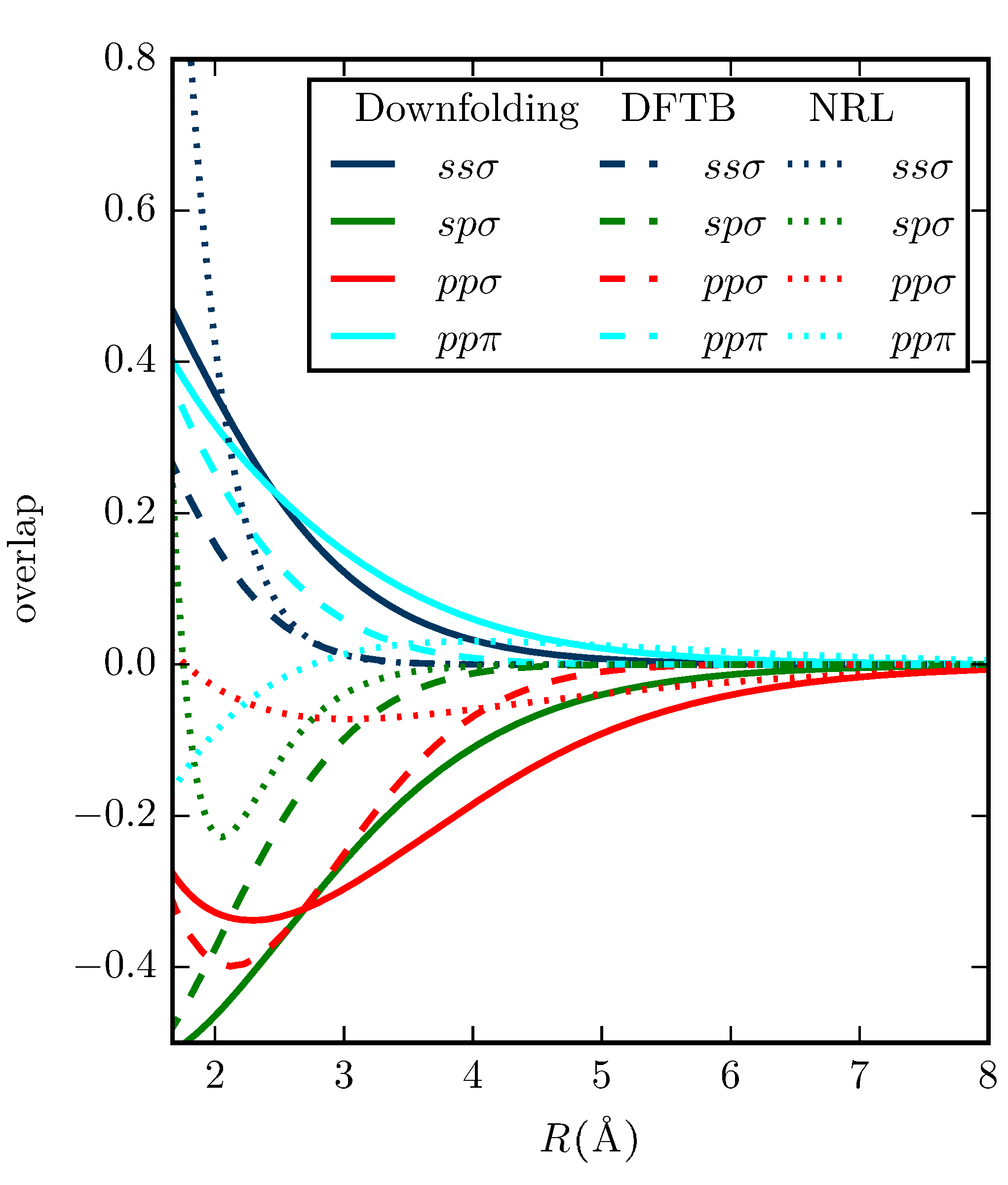}
\label{fig:comp_TB_S}
\caption{Comparison of TB matrix elements of Hamiltonian  $H$ (left) and overlap $S$ (right) of the Si-Si dimer obtained from downfolding, NRL-TB and DFTB. The sign of the $sp\sigma$ matrix element for DFTB was changed for easier comparison.}
\label{fig:comp_TB}
\end{figure*}

Here, we compare our TB parameterizations to NRL-TB and DFTB for the Si dimer.
The NRL-TB parameters for Si~\cite{papaconstantopoulos_mehl_erwin_pederson_1997} were chosen to reproduce both the band structure and the total energy of different bulk structures.
From the different parameterizations of DFTB~\cite{dftb} we choose the one that was optimized to experimental values for the band structure of bulk Si~\cite{7014245, 7182304}.
In Fig.~\ref{fig:comp_TB} we compare our non-orthogonal TB Hamiltonian and overlap matrix for the Si-Si dimer with the corresponding matrix elements for bulk Si in NRL-TB and DFTB. 
The Si-Si matrix elements in NRL-TB and DFTB decay faster with interatomic distance than the downfolded dimer matrix elements. This may be due to the contraction of the atomic orbitals in the bulk structures or a tight confinement potential. Our parameters of the Hamiltonian and even more so of the overlap matrix are closer to the DFTB than to the NRL-TB parameters.

\section{\label{sec:bulk} Transferability to bulk}

When atoms are grouped to form a bulk structure, we expect their valence states to contract and the overlap between two atoms to decrease due to screening contributions of neighboring bulk atoms. Therefore we must assume that the bond integrals obtained from dimers are not directly transferable to the bulk as they are too large in magnitude and range.

For a brief analysis of the prediction of the dimer bond integrals for bulk Si and Mo,
we limit the range of the bond integrals by multiplication with a cut-off function,
\begin{equation}
f_{\mathrm{cut}}(R) = \frac{1}{2} \left( \cos \left( \pi \left[ \frac{R - (R_{\mathrm{cut}} - d_{\mathrm{cut}})}{d_{\mathrm{cut}}}  \right] \right) \right),
\end{equation} 
with $R_{\mathrm{cut}} = 4 \text{\AA}$ for Si, $R_{\mathrm{cut}} = 6 \text{\AA}$ for Mo and $d_{\mathrm{cut}} = 0.5 \text{\AA}$ for both elements . 

\begin{figure*}[htb!]
\includegraphics[width=0.9\columnwidth]{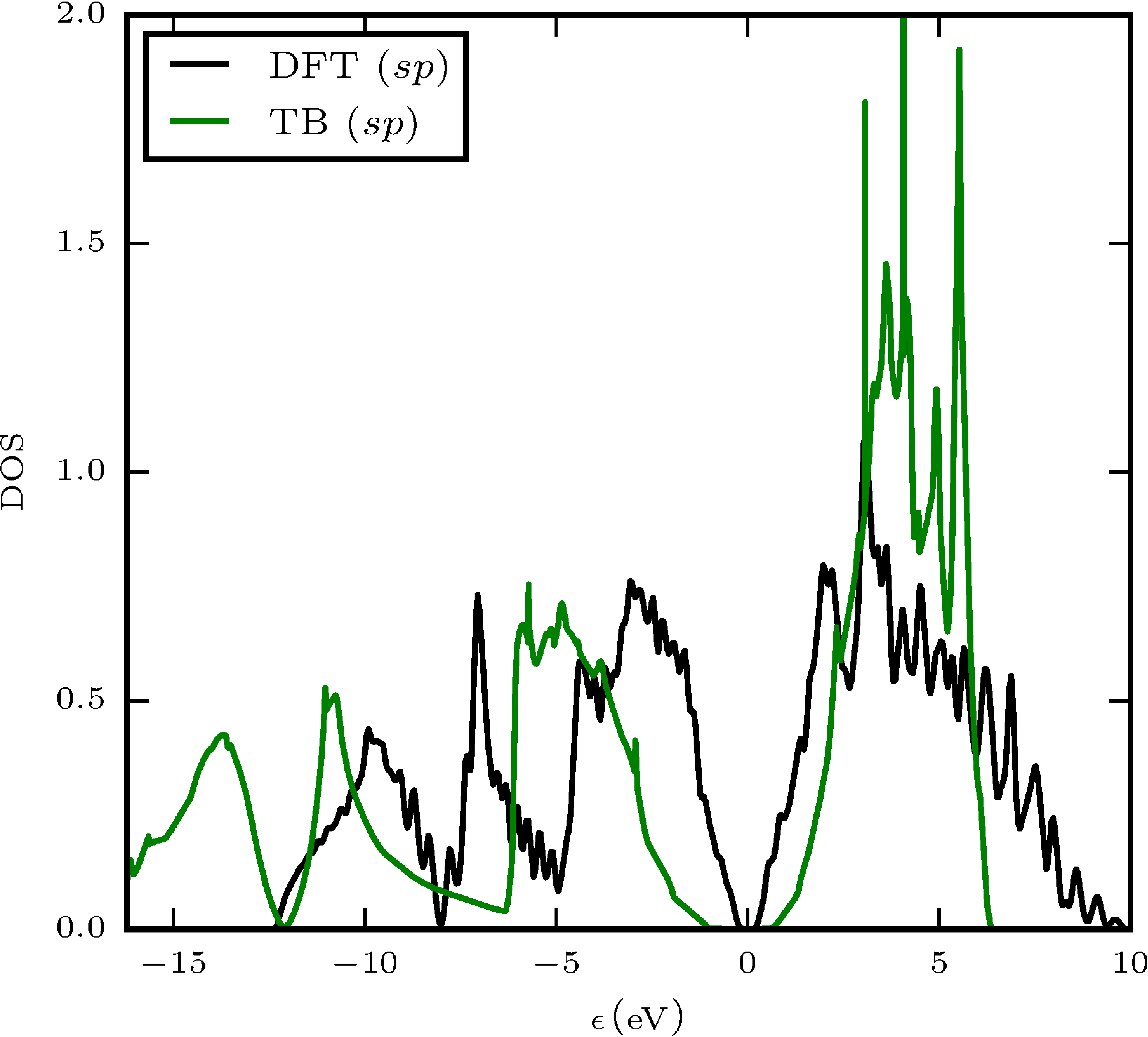}
\label{fig:Si_DOS_O}
\qquad
\includegraphics[width=0.9\columnwidth]{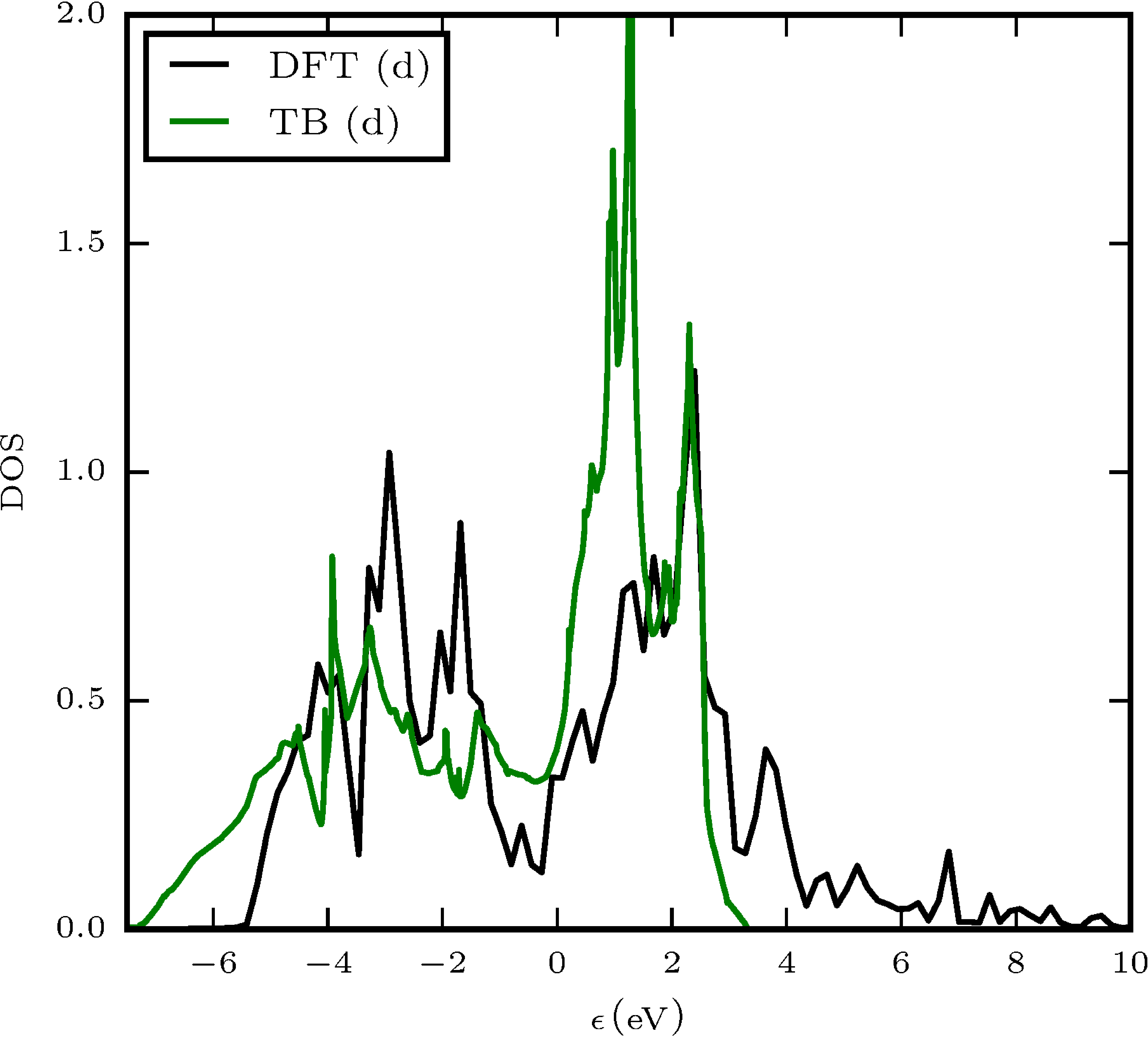}
\label{fig:Mo_DOS_O}
\caption{Electronic DOS (states/eV) of Si in the diamond structure (left) and Mo in the body-centered cubic structure (right) as obtained from DFT and from TB with the dimer TB bond parameters.}
\label{fig:DOS}
\end{figure*}

In Fig.~\ref{fig:DOS} we compare the electronic DOS obtained by TB as implemented in BOPfox~\cite{Hammerschmidt-19} and by self-consistent DFT using GPAW~\cite{PhysRevB.71.035109, 0953-8984-22-25-253202}.
The DOS obtained from our TB bond parameters for the Si dimer considerably overestimates the band width due to the neglect of screening contributions for the first neighbors. 
Nevertheless, the DOS is in good qualitative agreement with the results of Ref.~\cite{PhysRevB.91.054109} that used a projection of the self-consistent DFT eigenstates of bulk structures~\cite{PhysRevB.84.155119}. 
The TB model for Mo underestimates the width of the $d$-band but captures the bimodal character of the DOS that governs the structural stability of bcc Mo. 
The DOS can be further improved by fitting the parameters of the downfolded TB Hamiltonian to the electronic structure as it is done in the NRL-TB and DFTB models.

The TB bond parameters obtained in this work have already been used to construct models that capture phase transitions of bulk Ti~\cite{Ferrari-19}, the segregation of Re to partial dislocation in fcc Ni~\cite{Katnagallu-19} and the structural stability of different bulk Fe phases~\cite{Ladines-20}.
The respective refinements of the model parameters for the description of the relevant bulk phases required only moderate changes of the TB bond parameters for dimers obtained in this work.

\section{\label{sec:conclusion}Conclusions}

We parameterize TB bond parameters for nearly all combinations of elements of period 1 to 6 and group 3 to 18 of the periodic table. 
By downfolding the dimer DFT wave function in the Harris-Foulkes approximation to a minimal basis, we obtain the non-orthogonal TB Hamiltonian matrix, the overlap matrix and the L\"owdin-orthogonalized TB Hamiltonian matrix for 1711 homoatomic and heteroatomic dimers.

The TB eigenvalues compare well to their DFT reference over a wide range of interatomic distances. The TB matrix elements are smooth functions that are parameterized efficiently with only few exponential functions.

We demonstrate that the TB matrix elements follow intuitive chemical trends across the elements. By comparing  to well-known qualitative TB models, we rationalize and point out the limitations of the rectangular $d$-band model, a reduced TB model for $sp$ systems and canonical TB models for $sp$-valent and $d$-valent systems.

We briefly compare our parameterizations to NRL-TB and DFTB, a more detailed comparison requires taking into account the screening of the dimer bond integrals when they are immersed in the bulk.

The parameters for the 1711 dimers are provided in the Supplemental Material at Ref.~\cite{SuppMat} and may serve as the starting point for the parameterization of TB models with environmentally dependent matrix elements for transferability from free atoms to the bulk.

\begin{acknowledgments}
We wish to dedicate this paper to the memory of our coauthor, Prof. David Pettifor CBE FRS, who sadly passed away before the work was completed. J.J. acknowledges funding through the project ``Damage Tolerant Microstructures
in Steel`` by ThyssenKrupp Steel Europe AG and Benteler Steel/Tube GmbH. A.N.L., T.H., and R.D. acknowledge financial support by Deutsche Forschungsgemeinschaft (project C1 of collaborative research center SFB/TR 103). We are grateful to Nikita Medvedev for very useful comments on the manuscript.
\end{acknowledgments}

\appendix

\section{Block matrices of dimer Hamiltonian\label{app:BlockMatrices}}

We list the Hamiltonian matrix elements that are required in two-center approximation as characterized by the representations of the groups $D_{\infty h}$ and $C_{\infty v}$ that leave homoatomic and heteroatomic dimers invariant, respectively. The matrix elements of the $\sigma_0$, $\pi_{\pm 1}$ and $\delta_{\pm 2}$ blocks are given in Tab.~\ref{tab:submatrices} for the different combinations of valences of the two dimer atoms.
\begin{table*}[htb!]
\centering
\begin{tabular}{cccc}
\hline\hline
 valence&    $\sigma$&    $\pi$&    $\delta$\\ \hline

 $s$-$s$&
$\begin{pmatrix}
E_1^s&        ss\sigma\\
ss\sigma&    E_2^s\\
\end{pmatrix}$ &
&
\\ \hline

 $s$-$sp$&
$\begin{pmatrix}
E_1^s&        ss\sigma&    sp\sigma\\
ss\sigma&    E_2^s&        E_2^{s,p_0}\\
sp\sigma&    E_2^{s,p_0}&    E_2^{p_0}\\
\end{pmatrix}$ &
$\begin{pmatrix}
E_2^{p_{\pm1}}\\
\end{pmatrix}$ &
\\ \hline

 $s$-$sd$ &
$\begin{pmatrix}
E_1^s&        ss\sigma&    sd\sigma\\
ss\sigma&    E_2^s&        E_2^{s,d_0}\\
sd\sigma&    E_2^{s,d_0}&    E_2^{d_0}\\
\end{pmatrix}$ &
$\begin{pmatrix}
E_2^{p_{\pm1}}\\
\end{pmatrix}$ &
$\begin{pmatrix}
E_2^{d_{\pm2}}
\end{pmatrix}$ \\ \hline

 $sp$-$sp$&
$\begin{pmatrix}
E_1^s&        E_1^{s,p_0}&    ss\sigma&    sp\sigma\\
E_1^{s,p_0}&    E_1^{p_0}&    ps\sigma&    pp\sigma\\
ss\sigma&    ps\sigma    &    E_2^s&        E_2^{s,p_0}\\
sp\sigma&    pp\sigma&    E_2^{s,p_0}&    E_2^{p_0}\\
\end{pmatrix}$ &
$\begin{pmatrix}
E_1^{p_{\pm1}}&    pp\pi\\
pp\pi&            E_2^{p_{\pm1}}\\
\end{pmatrix}$ &
\\ \hline

 $sp$-$sd$ &
$\begin{pmatrix}
E_1^s&        E_1^{s,p_0}&    ss\sigma&    sd\sigma\\
E_1^{s,p_0}&    E_1^{p_0}&    ps\sigma&    pd\sigma\\
ss\sigma&    ps\sigma    &    E_2^s&        E_2^{s,d_0}\\
sd\sigma&    pd\sigma&    E_2^{s,d_0}&    E_2^{d_0}\\
\end{pmatrix}$ &
$\begin{pmatrix}
E_1^{p_{\pm1}}&    pd\pi\\
pd\pi&            E_2^{d_{\pm1}}\\
\end{pmatrix}$ &
$\begin{pmatrix}
E_2^{d_{\pm2}}
\end{pmatrix}$ \\  \hline

 $sd$-$sd$ &
$\begin{pmatrix}
E_1^s&        E_1^{s,d_0}&    ss\sigma&    sd\sigma\\
E_1^{s,d_0}&    E_1^{d_0}&    ds\sigma&    dd\sigma\\
ss\sigma&    ds\sigma    &    E_2^s&        E_2^{s,d_0}\\
sd\sigma&    dd\sigma&    E_2^{s,d_0}&    E_2^{d_0}\\
\end{pmatrix}$ &
$\begin{pmatrix}
E_1^{d_{\pm1}}&    dd\pi\\
dd\pi&            E_2^{d_{\pm1}}\\
\end{pmatrix}$ &
$\begin{pmatrix}
E_1^{d_{\pm2}}&    dd\delta\\
dd\delta&            E_2^{d_{\pm2}}\\
\end{pmatrix}$ \\
 \hline \hline
\end{tabular}
\caption{$\sigma$-, $\pi$- and $\delta$-block matrices of the dimer Hamiltonian matrix for different combinations of orbitals on the two dimer atoms.}
\label{tab:submatrices}
\end{table*}

\bibliography{mybibfile}

\end{document}